\documentclass[aps,twocolumn,pre,superscriptaddress]{revtex4-1}

\usepackage{xcolor,hyperref}
\hypersetup{
   colorlinks,
   linkcolor={blue!50!black},
   citecolor={blue!50!black},
   urlcolor={blue!80!black}
}

\usepackage{amsmath}
\usepackage{amssymb}
\usepackage{color}
\definecolor{mypink2}{RGB}{219, 48, 122}

\usepackage{graphicx}

\DeclareMathAlphabet{\mathitbf}{OML}{cmm}{b}{it}

\newcommand{\uv}{\mathitbf u}
\newcommand{\xv}{\mathitbf x}
\newcommand{\dv}{\mathitbf d}

\newcommand{\psiv}{\mathBold\psi}

\newcommand{\calBold}[1]{\mbox{\boldmath${\cal #1}$}}
\newcommand{\mathBold}[1]{\mbox{\boldmath$#1$}}
\newcommand{\dbar}{{\,\mathchar'26\mkern-12mu d}}

\begin{document}

\title{Mechanical disorder of sticky-sphere glasses. II. Thermo-mechanical inannealability}

\author{Karina Gonz\'alez-L\'opez}
\affiliation{Institute for Theoretical Physics, University of Amsterdam, Science Park 904, Amsterdam, Netherlands}
\author{Mahajan Shivam}
\affiliation{School of Physical and Mathematical Sciences, Nanyang Technological University, Singapore 637371, Singapore}
\author{Yuanjian Zheng}
\affiliation{School of Physical and Mathematical Sciences, Nanyang Technological University, Singapore 637371, Singapore}
\author{Massimo Pica Ciamarra}
\affiliation{School of Physical and Mathematical Sciences, Nanyang Technological University, Singapore 637371, Singapore}
\affiliation{CNR-SPIN, Dipartimento di Scienze Fisiche, Università di Napoli Federico II, I-80126 Napoli, Italy}
\author{Edan Lerner}
\email{e.lerner@uva.nl}
\affiliation{Institute for Theoretical Physics, University of Amsterdam, Science Park 904, Amsterdam, Netherlands}

\begin{abstract}
Many structural glasses feature static and dynamic mechanical properties that can depend strongly on glass formation history. The degree of universality of this history-dependence, and what it is possibly affected by, are largely unexplored. Here we show that the variability of elastic properties of simple computer glasses under thermal annealing depends strongly on the strength of attractive interactions between the glasses' constituent particles -- referred to here as glass `stickiness'. We find that in stickier glasses the stiffening of the shear modulus with thermal annealing is strongly suppressed, while the thermal-annealing-induced softening of the bulk modulus is enhanced. Our key finding is that the characteristic frequency and density per frequency of soft quasilocalized modes becomes effectively invariant to annealing in very sticky glasses, the latter are therefore deemed `thermo-mechanically inannealable'. The implications of our findings and future research directions are discussed. 
\end{abstract}

\maketitle

\section{introduction}
One intriguing feature of structural glasses is the strong history dependence of their mechanical properties~\cite{eran_jim_soft_matter_2013,falk_langer_stz,barrat_yielding_jcp_2004,Schuh_review_2007,Harmon_apl_2007,wang_review_2012,eran_fracture_toughness_pra_2016,Ozawa6656,MW_yielding_2018_pre,fsp,sri_annealing_2018,corrado_prl_2015,Yoshino_2018,Li2015}. It has been known since the work of Shi and Falk~\cite{falk_prl_2005} that plastic strain localization is enhanced, and stress overshoots are more pronounced, in glasses cooled at lower rates from a melt prior to their deformation~\cite{barrat_yielding_jcp_2004,Harmon_apl_2007,Ozawa6656,Yoshino_2018}. Other work has shown that deforming computer glasses under one loading geometry can significantly alter the subsequent responses of the material in the same and other loading geometries~\cite{Bauschinger_pre_2010,sylvain_prl_2020_Bauschinger}. Advances in experimental techniques that allow high control over the precise `fictive temperature' at which a metallic glass falls out of equilibrium~\cite{Eran_mechanical_glass_transition} were used to demonstrate that the notch fracture toughness of the same material can change by more than a factor of two, depending on its fictive temperature, i.e., on its preparation history. 

Methodological advances in computational glass physics have helped to shed considerable light on the aforementioned history dependence of glasses' mechanical properties. In particular, the optimization of polydisperse soft-sphere~\cite{LB_swap_prx} and other~\cite{LB_ultrastable_MGs} models with respect to their efficiency under Swap Monte Carlo dynamics has allowed the creation of glasses from very deeply supercooled liquids, and the systematic study of those glasses' mechanical properties as a function of their preparation history \cite{Ozawa6656,LB_modes_2019,boring_paper,pinching_pnas}. These studies and others have established that a generic feature of computer \cite{cge_paper} and laboratory glasses~\cite{Eran_mechanical_glass_transition} is that their shear modulus $G$ typically increases with deeper supercooling prior to glass formation; some observations report a total annealing-induced variation of $G$ of up to $\approx$ 60\% in 3D~\cite{pinching_pnas} (and $>70\%$ in 2D~\cite{corrado_cge_statistics_jcp_2020}). It has also been shown that the relative, sample-to-sample fluctuations of elastic moduli, shown recently to control long wavelength wave attenuation rates \cite{scattering_prl_2020}, decrease by more than a factor of 3 in well-annealed glasses~\cite{boring_paper}.

Another manifestation of thermal annealing on glasses' elasticity is seen in the energetic, statistical and structural properties of soft, quasilocalized modes (QLMs) \cite{SchoberOligschleger1996,modes_prl_2016,experimental_inannealability_AM_2016,protocol_prerc,inst_note,cge_paper,pinching_pnas,LB_modes_2019}. These low-energy excitations were shown to exist in any structural glass quenched from a melt \cite{modes_prl_2020}, and presumably play important roles in dynamic glassy phenomena such as wave attenuation \cite{scattering_jcp}, elasto-plasticity \cite{micromechanics2016}, aging dynamics \cite{Schober_correlate_modes_dynamics} and structural relaxation in equilibrium supercooled liquids \cite{widmer2008irreversible}. Evidence that a subset of these excitations may constitute the tunneling two-level systems, responsible for the anomalous thermodynamic and transport properties of glasses at cryogenic temperatures, has also been put forward \cite{SciPost2016}.

Several observations have been made that establish a connection between the degree of thermal annealing of glasses and their featured abundance of QLMs. To the best of our knowledge, the first observation was made by Schober and Oligschleger \cite{SchoberOligschleger1996}, who argued that the relative absence of soft, quasilocalized vibrational modes in a model glass is attributed to both their stiffening and overall depletion. This assertion was further discussed in Ref.~\cite{pinching_pnas}, where the total number density of QLMs was demonstrated to following a Boltzmann-like law with respect to the parent \emph{equilibrium} temperature $T_p$ from which the studied glasses were instantaneously quenched. In the same work it was shown that the characteristic frequency of QLMs increases by more than a factor of two due to strong thermal annealing.

What features of model structural glasses' interaction potentials control the degree of susceptibility of statistical-mechanical properties of a glass --- in particular its macro- and microscopic elastic properties, and mechanical disorder~\cite{footnote3} --- to thermal annealing? This is precisely the question we address in the present work, which is the second in a series of reports aimed at tracing out the effects of strong, attractive interactions between the constituent particles of a glass, on that glass's mechanical disorder. In the first paper \cite{sticky_spheres_part_1}, we aimed at avoiding thermal annealing effects on elasticity, in order to cleanly single-out the role of strong attractive interactions in determining glasses' elastic properties, mechanical disorder and stability. In order to compare between different computer glass models on the same footing, we followed Refs.~\cite{boring_paper,modes_prl_2020} and exploited the high-parent-temperature plateau of elastic properties featured by most computer glass models, including those considered in this work. 

Here we show that in a model computer glass in which the relative strength of pairwise attractive interactions --- referred to here as `glass stickiness' --- can be readily tuned \cite{potential_itamar_pre_2011}, the degree of thermal-annealing-induced \emph{variations} in glasses' elastic properties exhibits a strong dependence on glass stickiness. In particular, we find that the relative stiffening of the shear modulus in deeply annealed glasses decreases substantially upon increasing glass stickiness, while the relative softening of the bulk modulus increases. We further find that increasing glass stickiness leads to the indifference of the characteristic frequency scale associated with soft, quasilocalized excitations, of those excitations' characteristic size, and of their density per frequency, to thermal annealing. We refer to this surprising emergent indifference of mechanical properties to thermal annealing as \emph{thermo-mechanical inannealability} of glasses. 

\begin{figure}
\centering
  \includegraphics[width=0.95
  \linewidth]{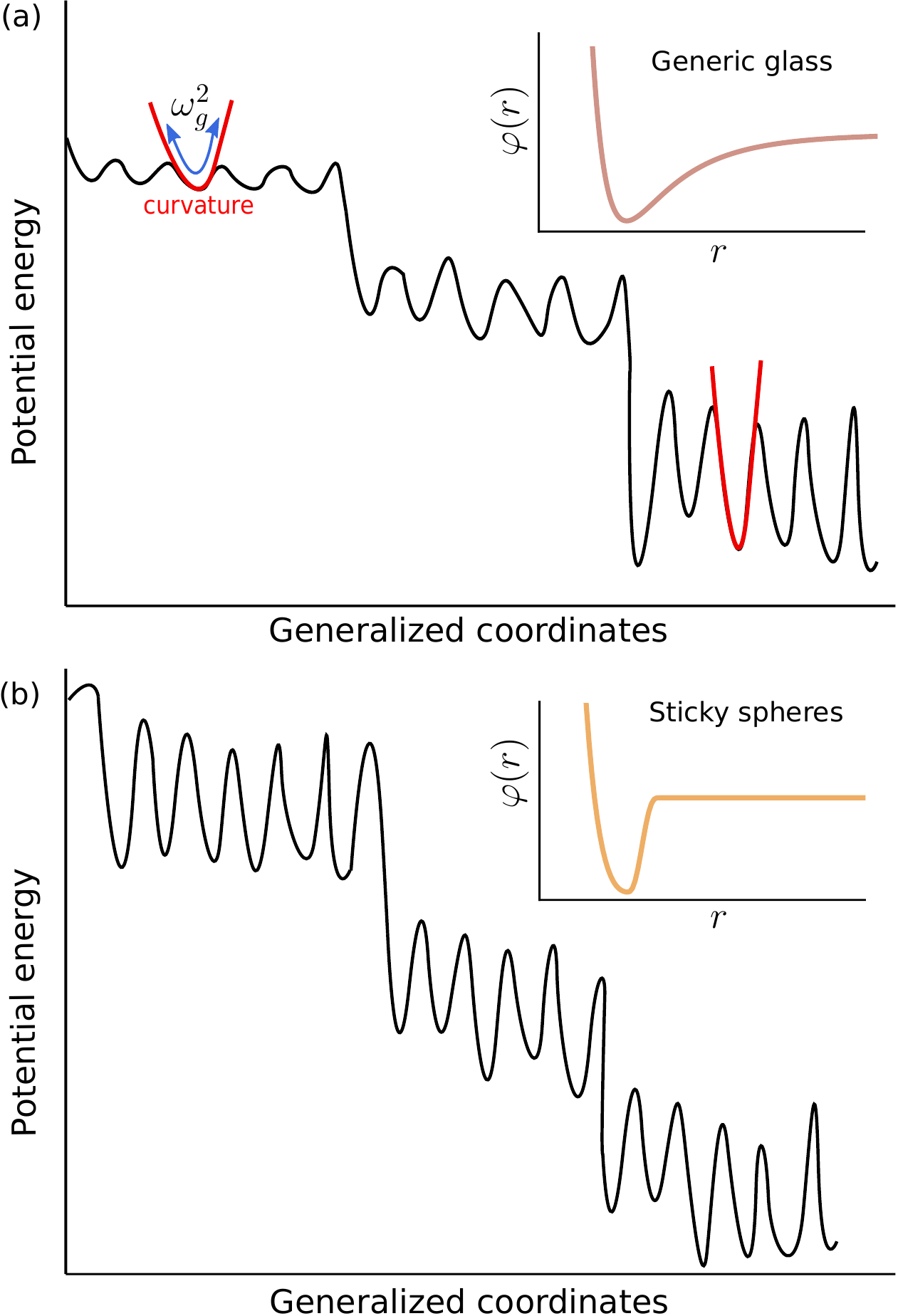}
\caption{\footnotesize Illustration of the concept of thermo-mechanical inannealability. The top panel depicts the potential energy landscape (PEL) of a generic glass. In generic glasses the characteristic frequencies associated with local minima \emph{increase} with decreasing energy. As we increase the relative strength of attractive forces and decrease their range (see insets) --- namely we make a `stickier' glass ---, the PEL is altered (see bottom panel): characteristic frequencies associated with local minima become largely \emph{independent} of the~energy.
\label{fig:thermomechanical_inannealability_cartoon}}
\end{figure}

The phenomenon of thermo-mechanical inannealability is illustrated within the potential energy landscape picture \cite{Goldstein1969} by the cartoon displayed in Fig.~\ref{fig:thermomechanical_inannealability_cartoon}. The potential landscape of a generic computer glass model --- e.g.~the extensively-studied Kob-Andersen Binary Lennard-Jones glass former~\cite{kablj} --- is illustrated in Fig.~\ref{fig:thermomechanical_inannealability_cartoon}a. In those systems, the characteristic frequency of the glass's soft nonphoninic quasilocalized modes --- represented by the characteristic \emph{curvature} of the landscape about its local minima, and denoted by $\omega_g^2$ in Fig.~\ref{fig:thermomechanical_inannealability_cartoon}a and in what follows  --- is small at high energies, and grows as lower energies are reached, see related discussions in Refs.~\cite{jeppe_review2006,wyart_vibrational_entropy,cge_paper,MW_cates_length_discussion_prl_2017}. 

In contrast with the behavior of generic glass-formers described above, the potential energy landscape of sticky-sphere glasses is substantially different. As illustrated in the bottom panel of Fig.~\ref{fig:thermomechanical_inannealability_cartoon}, the multidimensional landscape shows almost no change in its characteristic frequencies (curvatures about local minima) as one descends to lower energies. Below we will show that, despite that sticky spheres glass's potential energy features a \emph{larger} relative variation upon supercooling compared to glasses with weaker attractive interactions, many of these glasses' mechanical properties, including dimensionless quantifiers of mechanical disorder, are largely indifferent to descending deep down into the energy landscape.  

\vspace{0.5cm}
This work is structured as follows; in Sec.~\ref{sec:models} we introduce and motivate the choice of the model system employed in our study, and describe how different ensembles of glassy samples were created. In Sec.~\ref{sec:temperature_scale} we put forward a scheme that allows to extract a crossover temperature scale $T_{\mbox{\tiny co}}$ based on the potential energy per particle of our glass ensembles. The scale $T_{\mbox{\tiny co}}$ is then used to organize the rest of our data presentations. Sections \ref{sec:macro} and \ref{sec:meso} discuss the effect of particle stickiness on the thermal annealability of macroelastic and microelastic properties, respectively. We summarize our findings and discuss future research questions in Sec.~\ref{sec:summary_and_discussion}. The definitions of, and some explanations about, the physical observables considered in this work can be found in Appendix \ref{sec:appendix-defs}.


\section{Computer glass model and glass ensembles}
\label{sec:models}

\begin{table*}
\begin{ruledtabular}
\begin{tabular}{cccc}
$r_{c}$ & $T_{p}$ & $N$ & $n$\\
\hline
$1.1$& $6.00,4.00,3.00,2.00,1.50,1.20,1.00,0.88,0.80,0.76,0.73$ &3,000 & 9,200\\
$1.2$ & $4.00,2.60,1.80,1.30,1.00,0.91,0.85,0.80,0.77$ & 3,000 & 9,200\\
$1.3$ & $4.00,2.60,1.80,1.30,1.15,1.00,0.91,0.85,0.80,0.77$ & 3,000 & 9,200\\
$1.5$ & $0.76,0.72,0.69$ & 3,000 & 9,200\\
$1.5$ & $2.60,1.80,1.30,1.00,0.91,0.85,0.80$ & 10,000 & 3,000

\end{tabular}
\end{ruledtabular}
\caption{\label{sys_sizes_tp}
Equilibrium parent temperatures, for various cutoffs from the SS model. The system and ensemble size shown in the last two columns apply for all parent temperatures listed.}
\end{table*}

In this work we employ a 50:50 binary mixture of `large' and `small' particles of equal mass $m$ in three dimensions (3D) at fixed volume $V$. Pairs of particles interact via the Piecewise-Sticky-Spheres (PSS) pairwise potential, introduced first in Ref.~\cite{potential_itamar_pre_2011}, and also studied extensively in our companion paper \cite{sticky_spheres_part_1}. The equilibrium supercooled liquid dynamics of this model was very recently studied in Ref.~\cite{Massimo_supercooled_PRL}. The PSS model was chosen for this study since in Ref.~\cite{sticky_spheres_part_1} it featured the strongest variation of elastic properties as a function of its key control parameter, which is described next.

\begin{figure}[ht!]
\centering
  \includegraphics[width=1.0\linewidth]{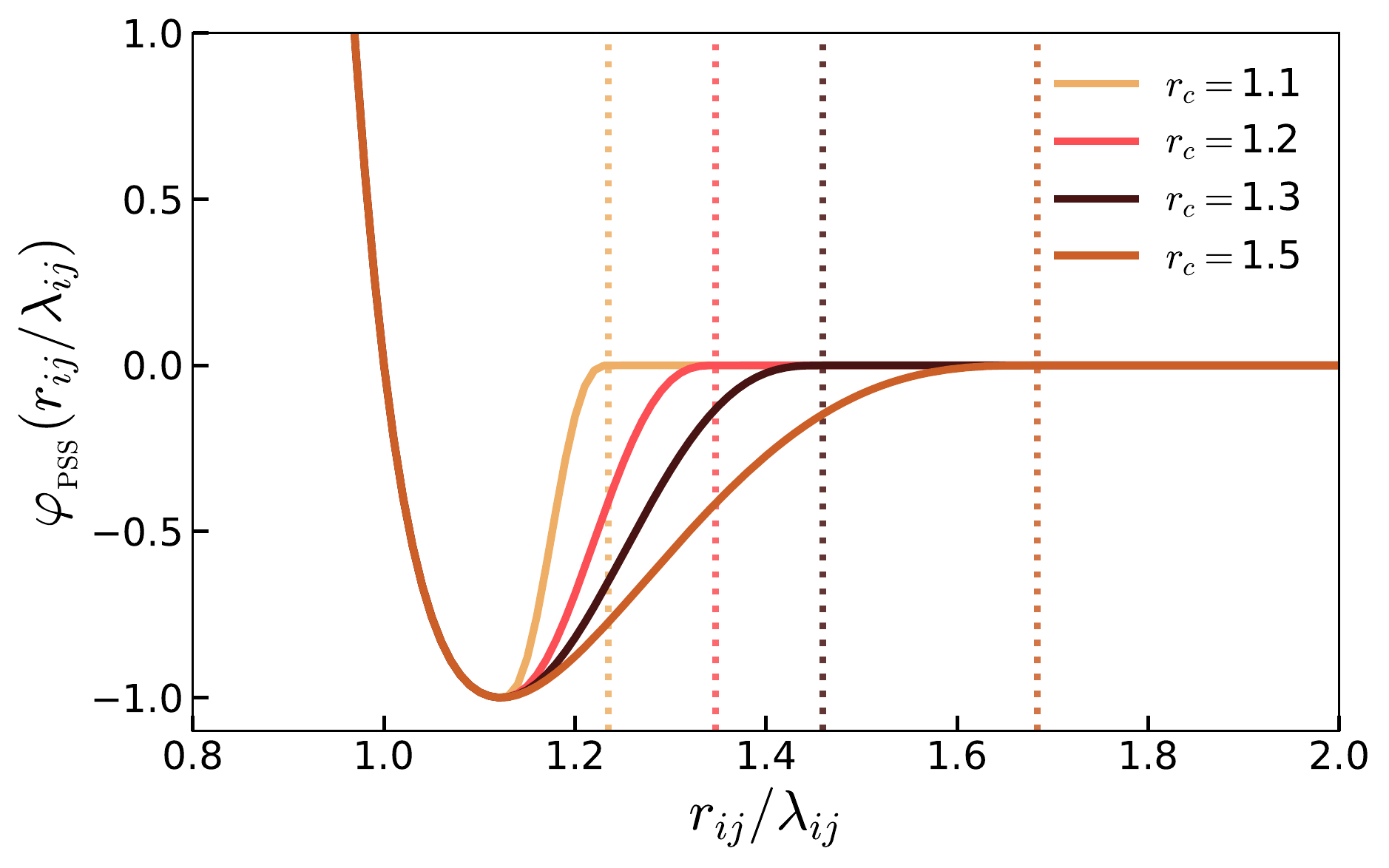}
\caption{\footnotesize The Piecewise Sticky Spheres (PSS) pairwise interaction potential employed in this work. The interaction cutoffs $r_c$ --- marked by the color-coded vertical dotted lines --- are expressed in terms of the dimensionless distance $x_{\mbox{\tiny min}}\!=\!2^{1/6}$ at which the canonical Lennard Jones potential attains a minimum.  \label{fig:potentials}}
\end{figure}

The PSS is a Lennard-Jones-like pairwise potential in which the repulsive part is identical to the canonical Lennard Jones (LJ) potential, but the attractive part is modified such that it and its first two derivatives with respect to interparticle distance $r_{ij}$ vanish continuously at a (dimensionless) cutoff distance $x_c$, the latter serving as a control parameter; the PSS pairwise potential reads
\begin{widetext}
\begin{equation}
    \varphi_{\mbox{\tiny PSS}}(r_{ij}) =
\left\{
\begin{array}{cc}
4\varepsilon \left[ \big(\frac{\lambda_{ij}}{r_{ij}}\big)^{12} - \big(\frac{\lambda_{ij}}{r_{ij}}\big) ^{6} \right],
     &  \frac{r_{ij}}{\lambda_{ij}}< x_{\mbox{\tiny min}}  \\
\varepsilon \left[a\big(\frac{\lambda_{ij}}{r_{ij}}\big)^{12} -b\big(\frac{\lambda_{ij}}{r_{ij}}\big)^{6} + \sum\limits_{\ell=0} ^{3}  c_{\mbox{\tiny $2\ell$}} \big(\frac{r_{ij}}{\lambda_{ij}}\big)^{2\ell} \right] , & x_{\mbox{\tiny min}}\le \frac{r_{ij}}{\lambda_{ij}}< x_c\\
0\,,  & \frac{r_{ij}}{\lambda_{ij}}\geq x_c
\end{array}
\right. ,
 \label{eq:wideeq}
\end{equation}
\end{widetext}
where $\varepsilon$ is a microscopic energy scale, $x_{\mbox{\tiny min}},x_c$ are the (dimensionless) locations of the minimum of the LJ potential and modified cutoff, respectively, and the length parameters $\lambda_{ij}$ are expressed in terms of the `small-small' interaction length $\lambda_{\mbox{\tiny small}}^{\mbox{\tiny small}}$, with $\lambda_{\mbox{\tiny small}}^{\mbox{\tiny large}}\!=\!1.18\lambda_{\mbox{\tiny small}}^{\mbox{\tiny small}}$ and $\lambda_{\mbox{\tiny large}}^{\mbox{\tiny large}}\!=\!1.4\lambda_{\mbox{\tiny small}}^{\mbox{\tiny small}}$. The coefficients $a,b,\{c_{\mbox{\tiny $2\ell$}}\}$ can be found in Ref.~\cite{sticky_spheres_part_1}; they are chosen such that the attractive and repulsive parts of the potential and two derivatives are continuous at $x_{\mbox{\tiny min}}$ and at $x_c$. The pairwise potential $\varphi_{\mbox{\tiny PSS}}$ is plotted in Fig.~\ref{fig:potentials}. In what follows, we express the dimensionless cutoff $x_c$ of $\varphi_{\mbox{\tiny PSS}}$ in terms of $x_{\mbox{\tiny min}}\!=\!2^{1/6}$ by defining $r_c\!\equiv\! x_c/x_{\mbox{\tiny min}}$, for simplicity. $r_c$~serves as one of the two key control parameters in our investigation.

We build ensembles of glassy samples at fixed number density $N/V\!=\!0.60(\lambda_{\mbox{\tiny small}}^{\mbox{\tiny small}})^{-3}$, using $r_c\!=\!1.1,1.2,1.3$, and 1.5; glass configurations were initially equilibrated at various parent temperatures $T_p$ --- the second key control parameter in our investigation --- for which the liquids' equilibrium relaxation times (estimated via the stress autocorrelations, see Fig.~\ref{fig:stress_correlations} in Appendix~\ref{sec:appendix-defs}) vary over approximately 4 orders of magnitude, for each $r_c$. Details about the equilibrium parent temperatures, system and ensembles sizes appear in Table~\ref{sys_sizes_tp}. We have checked and found no signs of crystallization. In what follows, we report all lengths in terms of the characteristic interparticle distance $a_0\!\equiv\!(V/N)^{1/3}$, and all frequencies in terms of $\omega_0\!\equiv\! c_s/a_0$, where the speed of shear waves is defined as $c_s\!\equiv\!\sqrt{G/\rho}$ with $\rho\!\equiv\!mN/V$ denoting the mass density, and notice that $\omega_0$, $c_s$ and $G$ are all $r_c$- and $T_p$-dependent.

Importantly, we note that the thermal-annealing-induced variation percentages of the observables reported in what follows depend on the depth of supercooling of glasses' ancestral equilibrium configurations; here, our different glass models (pertaining to different cutoffs $r_c$) were all supercooled roughly evenly, at least in terms of their respective equilibrium liquid dynamics as shown in Fig.~\ref{fig:stress_correlations} in Appendix~\ref{sec:appendix-defs}. This roughly-even depth of supercooling across models allows us to meaningfully compare thermal-annealing-induced \emph{relative} variations of observables, across the entire parent temperature range of each glass model, and across ensembles of different glass models.

\section{Extracting a crossover temperature scale}
\label{sec:temperature_scale}

In the companion paper \cite{sticky_spheres_part_1} of this series, we prepared glassy samples by instantaneously quenching high temperature equilibrium liquid states; the equilibrium parent temperatures were chosen to be at least a factor of~4 higher than the computer glass transition temperature $T_g$, defined here as the temperature at which the structural relaxation time $\sim10^4\tau_\star$, with $\tau_\star$ representing a characteristic vibrational time scale. As we shall see below, and as previously observed \cite{Sastry1998,SASTRY1999301,SASTRY2002267,boring_paper}, elastic properties and various dimensionless quantifiers of mechanical disorder exhibit a plateau above some crossover temperature, denoted in what follows as $T_{\mbox{\tiny co}}$. An equal-footing comparison of elastic properties is possible by considering parent temperatures $T_p$ much larger than the crossover temperature $T_{\mbox{\tiny co}}$. 

How should the crossover temperature $T_{\mbox{\tiny co}}$ --- with respect to which we compare different glasses made by quenching states equilibrated at various parent temperatures $T_p$ --- be defined? Defining a physically-relevant temperature scale in supercooled liquids is not a trivial task, see e.g.~discussions in Refs.~\cite{Reichman_onset_pre_2004,sri_crossover_jcp_2017}. Here we introduce a simple, broadly applicable and evidently useful scheme to define $T_{\mbox{\tiny co}}$, which is based on the $T_p$-dependence of the ensemble average energy per particle $u(T_p)\!\equiv\!U(T_p)/N$ of our computer glasses. The raw data for $u(T_p)$, expressed in terms of simulational units, are shown in Fig.~\ref{fig:raw_potential_energy} in Appendix~\ref{sec:potential_energy_raw_data}.

\begin{figure}[h!]
\centering
  \includegraphics[width=1.00\linewidth]{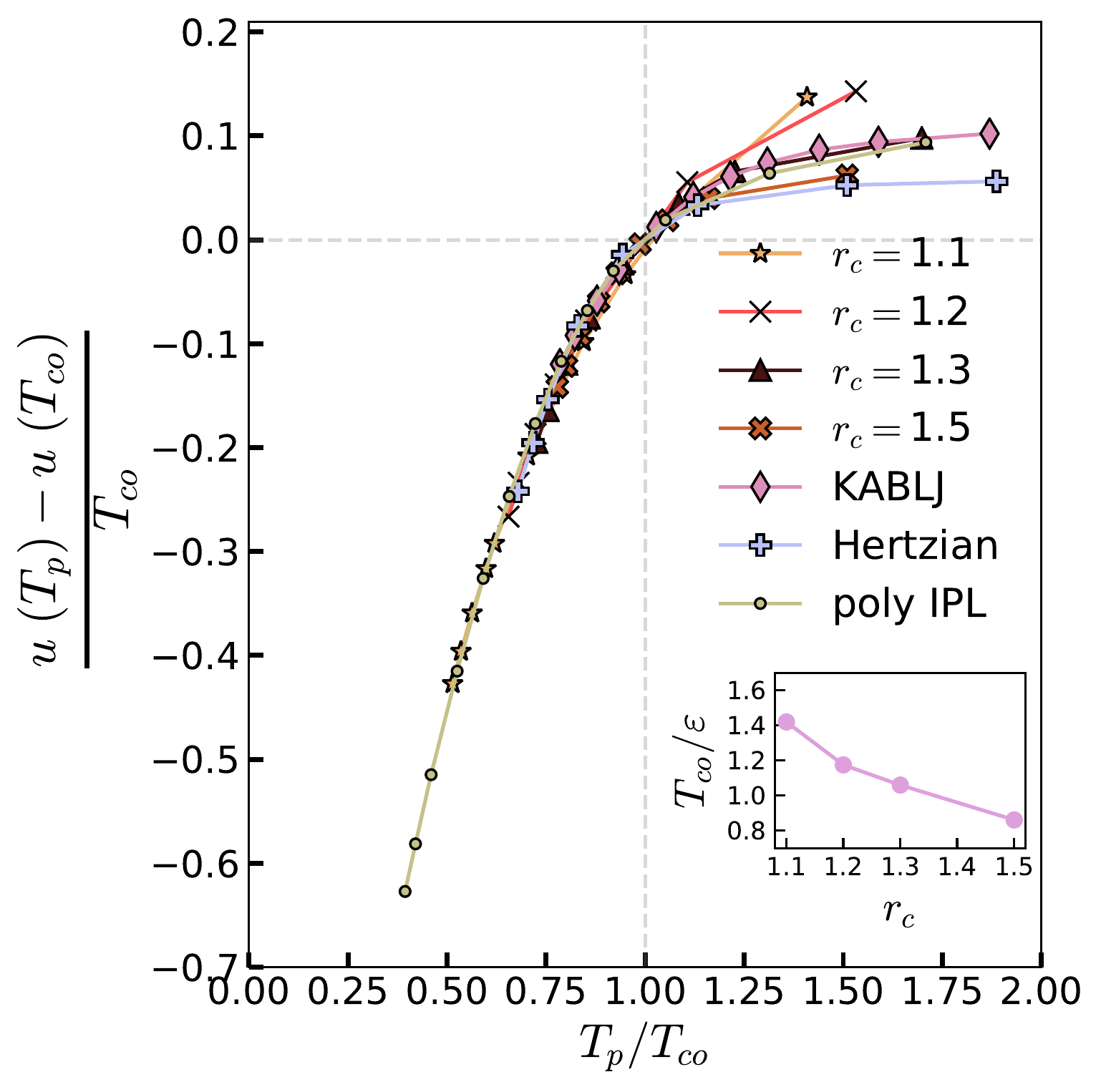}
\caption{\footnotesize Glass potential energy per particle $u(T_p)$, shifted by the interpolated $u(T_{\mbox{\tiny co}})$, and rescaled by the crossover temperature $T_{\mbox{\tiny co}}$, see text for discussion. The inset shows the extracted crossover temperatures $T_{\mbox{\tiny co}}$ for the computer glasses studied in this work.  \label{fig:collapse_energies}}
\end{figure}

\begin{figure*}[!ht]
\centering
  \includegraphics[width=1.0\linewidth]{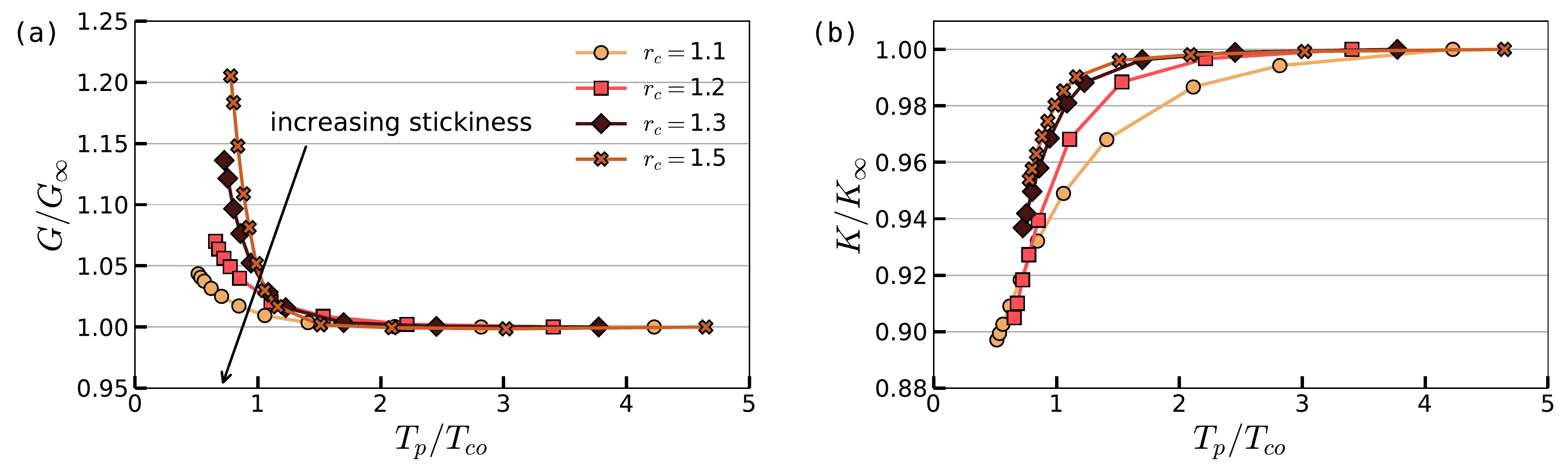}
\caption{\footnotesize (a) Sample-to-sample average shear modulus $G$, rescaled by its high temperature plateau value $G_{\infty}$, and plotted against $T_p/T_{\mbox{\tiny co}}$. (b) same as (a) for the sample-to-sample mean bulk modulus $K$, see text for discussion.
\label{fig:g_over_k_dimless_tps}}
\end{figure*}

The scheme is illustrated in Fig.~\ref{fig:collapse_energies}; we show $(u(T_p)-u(T_{\mbox{\tiny co}}))/T_{\mbox{\tiny co}}$ vs.~$T_p/T_{\mbox{\tiny co}}$, namely the difference between the inherent state energy per particle $u(T_p)$, and the linearly interpolated energy per particle $u(T_{\mbox{\tiny co}})$, rescaled by the crossover temperature $T_{\mbox{\tiny co}}$, and plotted against the rescaled parent temperature $T_p/T_{\mbox{\tiny co}}$. This is done by choosing the crossover temperatures $T_{\mbox{\tiny co}}$ for each model, such that the collapse across all models is optimal for $T_p\!<\! T_{\mbox{\tiny co}}$. The quasiuniversal form of the shifted and rescaled $u(T_p)$ is also shown in Fig.~\ref{fig:collapse_energies} to be followed by 3 additional computer glass models: the canonical Kob-Andersen Binary Lennard-Jones model \cite{kablj}, a Hertzian soft-spheres glass \cite{modes_prl_2020}, and a polydisperse, inverse power law glass \cite{boring_paper,pinching_pnas}, whose respective extracted crossover temperatures $T_{\mbox{\tiny co}}$ can be found in~\cite{footnote}.

Interestingly, despite a good collapse of $u(T_p)$ at $T_p\!<\!T_{\mbox{\tiny co}}$ for several different computer glass models, there seems to be no obvious connection between the extracted crossover temperatures $T_{\mbox{\tiny co}}$, and the supercooled dynamics of the parent equilibrium states from which the glasses were quenched. This is to say that, for example, the dimensionless structural relaxation time of the $r_c\!=\!1.1$, $T_p/T_{\mbox{\tiny co}}\!\approx\!0.52$ liquids is roughly equal to that of the $r_c\!=\!1.5$, $T_p/T_{\mbox{\tiny co}}\!\approx\!0.78$ liquids (see lowest-$T_p$ correlation functions in Fig.~\ref{fig:stress_correlations}a,d), implying that the dimensionless relaxation time is \emph{not} a universal function of $T_p/T_{\mbox{\tiny co}}$. Consequently, the dimensionless glass transition temperature $T_g/T_{\mbox{\tiny co}}$ is not expected to be universal either.

Another interesting observation, discussed further below, is that even though the crossover temperature $T_{\mbox{\tiny co}}$ is a decreasing function of the interaction potential cutoff $r_c$, the absolute variation of the rescaled and shifted glass energy $\big(u(T_p)\!-\!u(T_{\mbox{\tiny co}})\big)/T_{\mbox{\tiny co}}$ for $T_p\!<\!T_{\mbox{\tiny co}}$ appears to be larger for the stickier glasses, with the smaller cutoffs. This is to say that the largest variation of glass energy per particle --- in terms of $T_{\mbox{\tiny co}}$ --- is seen for the $r_c\!=\!1.1$ glasses, in contrast with the $T_p$-dependence of many of those glasses' \emph{elastic} properties, as shown in what follows.

An obvious limitation of the scheme described above is that it only allows the extraction of crossover temperatures if $u(T_p)$ is available for a few computer glass models, that all follow the same  quasiuniversal form below $T_p$, which is not \emph{a priori} known. However, a very close estimate of the extracted $T_{\mbox{\tiny co}}$ --- as seen in the inset of Fig.~\ref{fig:collapse_energies} --- can be obtained by analyzing the $T_p$-dependent elastic moduli of a single computer glass model, as explained and demonstrated in Appendix~\ref{sec:alternative_Tco_appendix}.

\section{Effect of Thermal annealing on macroelasticity}
\label{sec:macro}

Having established how to extract a crossover temperature scale $T_{\mbox{\tiny co}}$ for our different $r_c$-ensembles, we next review our measurements of macroelastic observables, namely elastic moduli. Precise definitions of the studied observables can be found in Appendix~\ref{sec:appendix-defs}.


Elastic moduli of glasses are known to depend on the equilibrium parent temperature $T_p$ from which those glasses were quenched \cite{eran_fracture_toughness_pra_2016,cge_paper,boring_paper,corrado_cge_statistics_jcp_2020}. Here we assess the degree of this dependence under variations of the relative strength of attractions (glass `stickiness'), tuned in turn by varying the interaction cutoff $r_c$ of our sticky-sphere glasses, as explained in Fig.~\ref{fig:potentials}. In Fig.~\ref{fig:g_over_k_dimless_tps}a we plot the sample-to-sample average athermal shear modulus $G$, rescaled by its high-$T_p$ plateau, denoted as $G_\infty$. We see that increasing glass stickiness leads to the suppression of the relative thermal-annealing-induced variation in $G$: for the $r_c\!=\!1.5$ glasses $G$ increases by slightly more than 20\%, whereas the $r_c\!=\!1.1$ glasses feature a much milder variation, of slightly less than 5\% -- a factor of more than 4 smaller relative thermal-annealing-induced variation compared to the $r_c\!=\!1.5$ glasses. This increasing indifference to thermal annealing by increasing glass stickiness is the first example of thermo-mechanical inannealability presented in this work. 

Interestingly, the ensemble-average athermal bulk modulus $K(T_p)$, plotted in Fig.~\ref{fig:g_over_k_dimless_tps}b after rescaling by its high-$T_p$ limit $K_\infty$, shows two opposite trends compared to $G(T_p)$; first, $K(T_p)$ is a decreasing function of the parent temperature $T_p$, whereas $G(T_p)$ is an increasing function of $T_p$. The observed decrease of $K$ with annealing appears to be a common feature of some simple glass models \cite{boring_paper,LB_modes_2019}, but not of laboratory glasses \cite{wang_review_2012,experimental_inannealability_AM_2016} that are typically annealed at constant pressure. Second, the largest relative decrease in $K(T_p)$ --- of roughly 10\% --- is featured by the stickiest amongst our glasses: the $r_c\!=\!1.1$ ensemble, while the largest relative increase in $G(T_p)$ is featured by the $r_c\!=\!1.5$ glasses.

\begin{figure}[ht!]
\centering
  \includegraphics[width=1.0\linewidth]{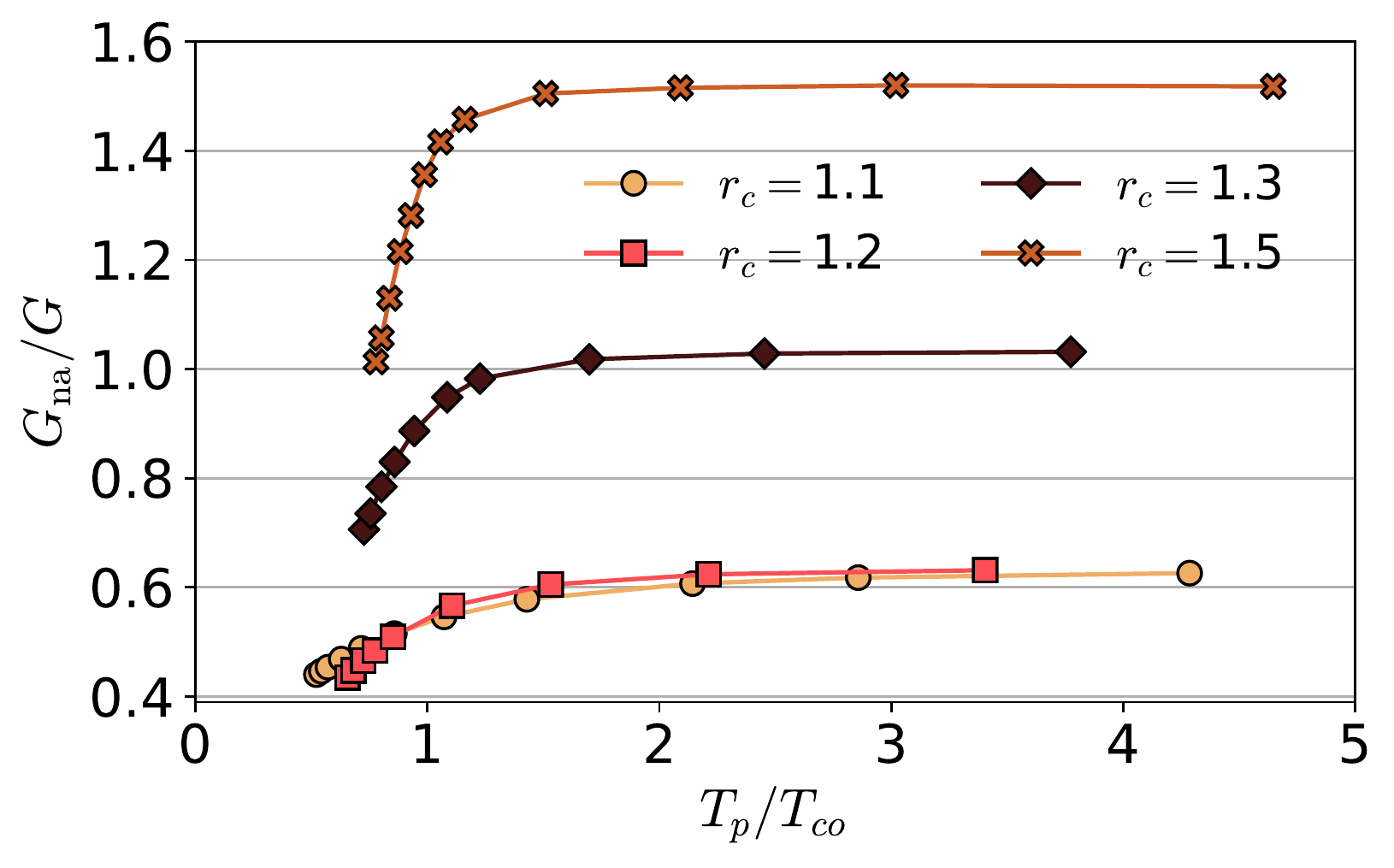}
\caption{\footnotesize The shear modulus $G$ can be decomposed into an affine part $G\!-\!G_{\rm{na}}$, and a nonaffine part $G_{\rm{na}}$ that features an explicit dependence on the vibrational spectrum of a glass, see precise definitions in Appendix~\ref{sec:appendix-defs}. Here we show the fraction $G_{\rm{na}}/G$ vs.~parent temperature $T_{\rm p}$ for all $r_c$-ensembles, see text for discussion.
\label{fig:non-affine}}
\end{figure}

Fig.~\ref{fig:non-affine} shows the fraction of the nonaffine term $G_{\rm{na}}/G$ from the total shear modulus $G$ (see definitions in Appendix~\ref{sec:appendix-defs}, and related work in the context of the unjamming transition in~\cite{Silbert_pre_2016_jamming}). As seen for many mechanical observables, also in this case we find that $G_{\rm{na}}/G$ features a high-$T_{\rm p}$ plateau, and a downwards dip at roughly $T_{\rm co}$. Interestingly, the high-$T_p$ plateau of $G_{\rm{na}}/G$ depends strongly on glass stickiness, varying by more than a factor of 2 across the entire $r_c$-range considered. At the same time, the relative $T_p$-induced variation of $G_{\rm{na}}/G$ is similar across the different studied degrees of glass stickiness, ranging roughly between 40\% and 50\%.

\begin{figure}[ht!]
\centering
  \includegraphics[width=1.0\linewidth]{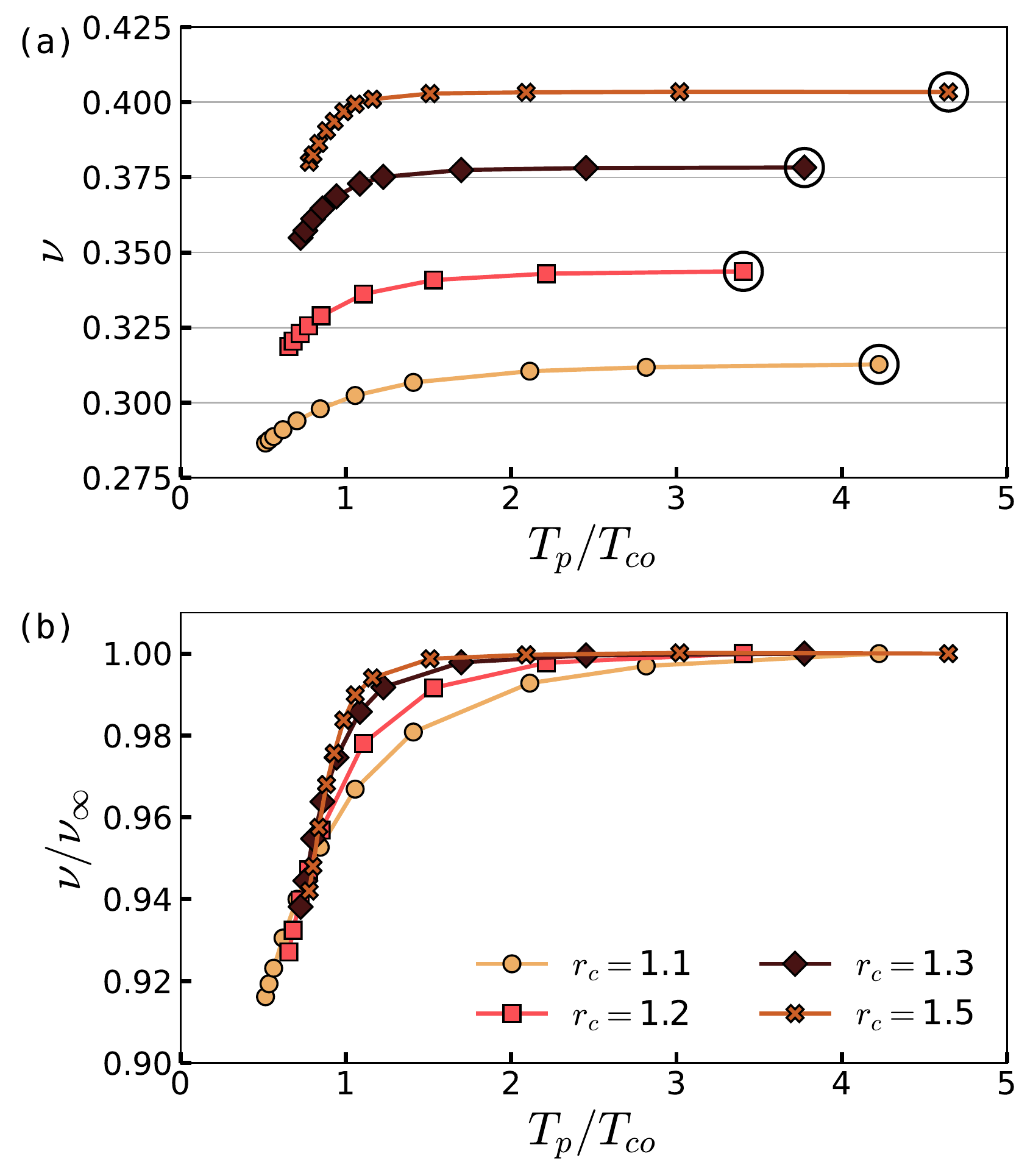}
\caption{\footnotesize (a) The Poisson's ratio $\nu$ is plotted against the parent temperature $T_p$, for all $r_c$-ensembles. The circled data points represent $\nu_\infty$, used for rescaling $\nu$ as seen in panel (b).
\label{fig:poissons_ratio}}
\end{figure}

Finally, we show in Fig.~\ref{fig:poissons_ratio}a the sample-to-sample average Poisson's ratio $\nu\!\equiv\!(3-2G/K)/(6+2G/K)$, plotted against $T_p$, for all $r_c$-ensembles. We find that while the typical $\nu$ values depend quite significantly on $r_c$ (as we have also shown in Ref.~\cite{sticky_spheres_part_1}), the relative variation of $\nu$ across the entire sampled $T_p$ range, as seen in Fig.~\ref{fig:poissons_ratio}b, does not differ much between the highest and lowest $r_c$-ensembles. Clearly, the rescaled curves $\nu/\nu_\infty$ do not collapse when plotted against $T_p/T_{\mbox{\tiny co}}$.

\section{Effect of Thermal annealing on microscopic elasticity}
\label{sec:meso}

The vibrational density of states (vDOS) associated with nonphononic soft quasilocalized modes in structural glasses has been long ago predicted \cite{soft_potential_model_1991,Gurevich2003} and recently shown \cite{modes_prl_2016} to universally follow a quartic law, namely
\begin{equation}\label{eq:vdos}
        {\cal D}(\omega)\!=\!A_{g}\omega^{4}\,,
\end{equation}
independent of spatial dimension \cite{modes_prl_2018}, glass formation protocol \cite{cge_paper,LB_modes_2019,pinching_pnas}, or form of microscopic interactions~\cite{modes_prl_2020}. The prefactor $A_g$ has dimensions of [frequency]$^{-5}$, and its physical essence has been discussed at length in Refs.~\cite{cge_paper,pinching_pnas}. In those works it is asserted that $A_g(T_p)\!\sim\!{\cal N}(T_p)\,\omega_g^{-5}(T_p)$, where ${\cal N}(T_p)$ and $\omega_g(T_p)$ represent the parent-temperature dependent density per particle and characteristic frequency of soft quasilocalized modes (QLMs), respectively.

Previous investigations have shown that thermal annealing of computer glasses can affect the statistical, energetic and structural properties of their embedded soft QLMs \cite{SchoberOligschleger1996,modes_prl_2016,protocol_prerc,cge_paper,boring_paper,LB_modes_2019,pinching_pnas}. Here we investigate how the susceptibility of those aforementioned properties to thermal annealing -- changes by varying glass stickiness.

\begin{figure}[!h]
\centering
  \includegraphics[width=1.0\linewidth]{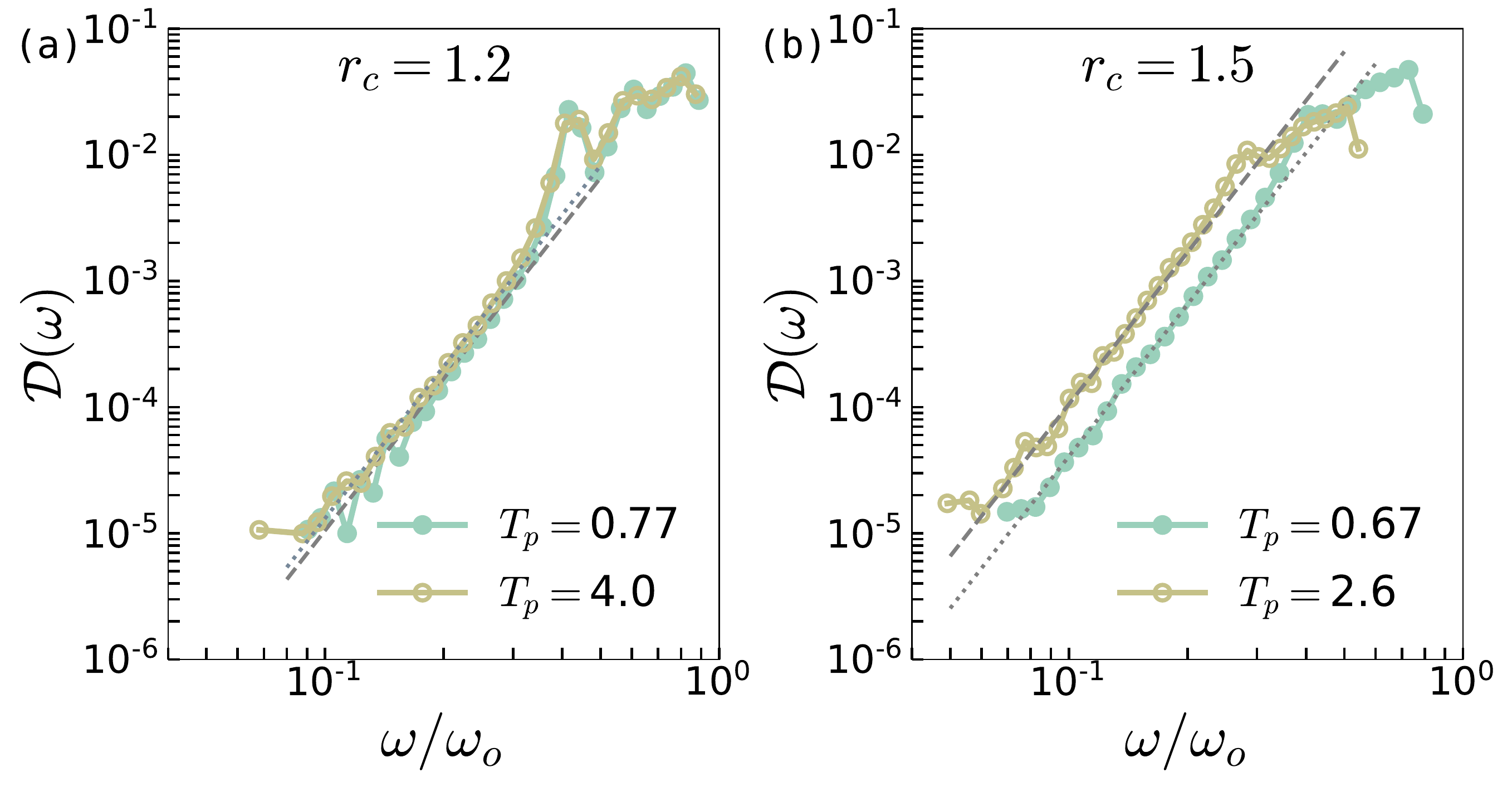}
\caption{ The vDOS ${\cal D}(\omega)$, calculated for low and high parent temperatures $T_p$, in our $r_c\!=\!1.2$ systems (left) and for our $r_c\!=\!1.5$ systems (right). We find that the prefactor $A_g$ of the $\omega^4$ scaling depends very weakly on $T_p$ for moderate particle stickiness ($r_c\!=\!1.2$), and shows a much more pronounced dependence on thermal annealing for weak particle stickiness ($r_c\!=\!1.5$).
\label{fig:ag_mesurement}}
\end{figure}

\subsection{Density per frequency of QLMs} \label{sec:vdos_and_pref}

We first study the thermal-annealing susceptibility of the prefactor $A_g(T_p)$, as measured in our different-$r_c$ glass-ensembles. In Fig.~\ref{fig:ag_mesurement} we show examples of the low-frequency vDOS of our computer glasses. In particular, we plot ${\cal D}(\omega)$ against $\omega/\omega_0$, for the highest and lowest parent temperatures, and for two cutoffs: $r_c\!=\!1.2$ (Fig.~\ref{fig:ag_mesurement}a) and $r_c\!=\!1.5$ (Fig.~\ref{fig:ag_mesurement}b). We recall that $\omega_0\!\equiv\!c_s/a_0$, where $c_s$ is the speed of shear waves, and $a_0$ an interparticle distance. Superimposed dashed and dotted lines are fits to Eq.~\ref{eq:vdos}, which demonstrate how $A_g$ is extracted from the vDOS data.

\begin{figure}[h!]
\centering
  \includegraphics[width=0.95\linewidth]{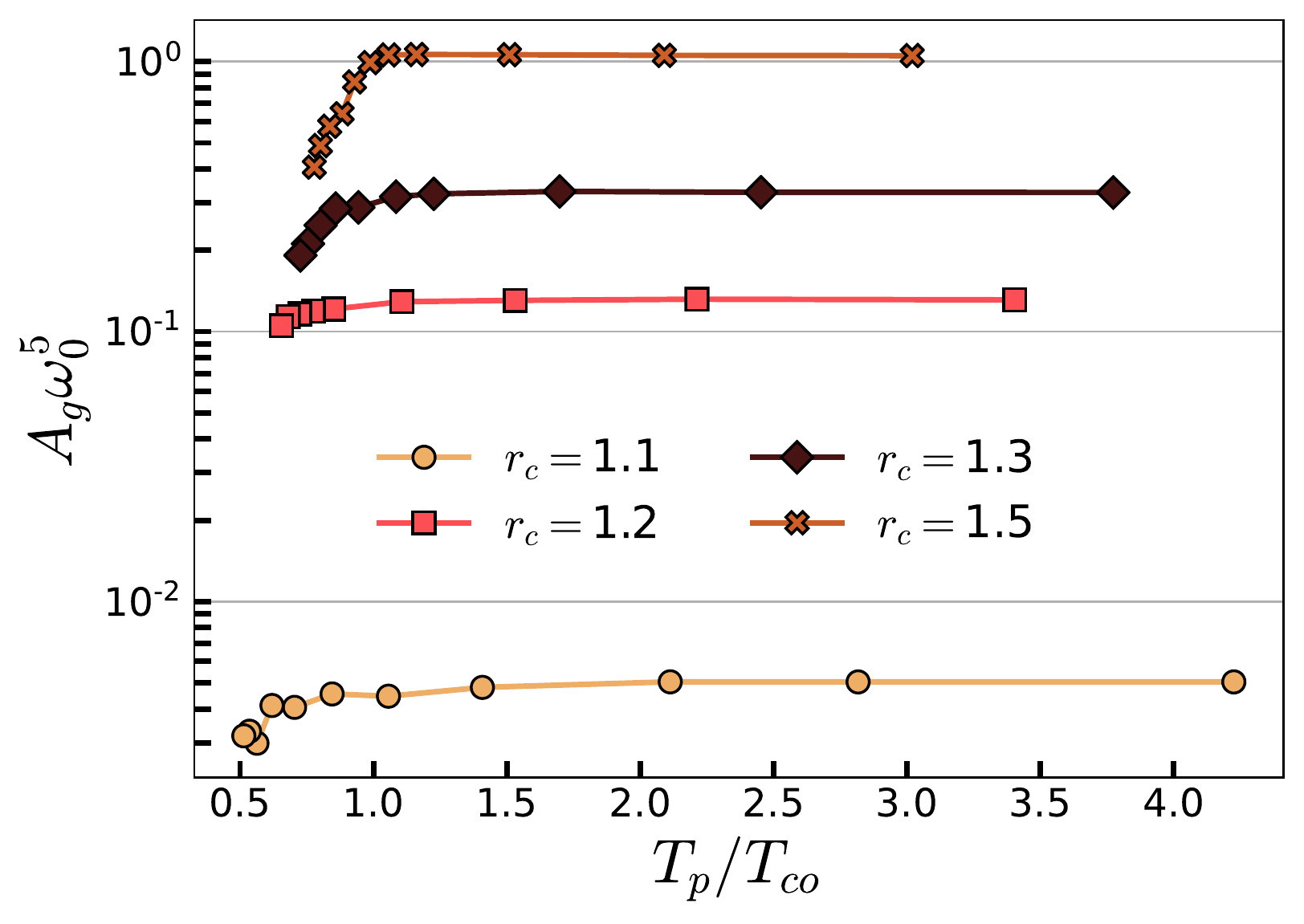}
\caption{\footnotesize Prefactors $A_g$ of the universal nonphononic vDOS ${\cal D}(\omega)\!=\!A_g\omega^4$, made dimensionless by scaling by $\omega_0^5$, measured across and plotted against parent temperatures $T_p$, for our different $r_c$-ensembles.
\label{fig:prefactor}}
\end{figure}

Comparing the vDOS of the $r_c\!=\!1.2$ and the $r_c\!=\!1.5$ glasses, we clearly see that in the former case $A_g$ is nearly independent of $T_p$, while in the latter case a measurable difference is seen between the high and low $T_p$ data. A more comprehensive presentation of the dependence of $A_g$ on $T_p$ is presented in Fig.~\ref{fig:prefactor}, where we plot $A_g$ against $T_p$ for all $r_c$'s. In the companion paper \cite{sticky_spheres_part_1} we showed that $A_g$ is very sensitive to increasing glass stickiness; that sensitivity can be seen here too by examining the large $T_p$ plateaus of $A_g$ for different $r_c$'s.

\begin{figure*}
\centering
  \includegraphics[width=1.0
  \linewidth]{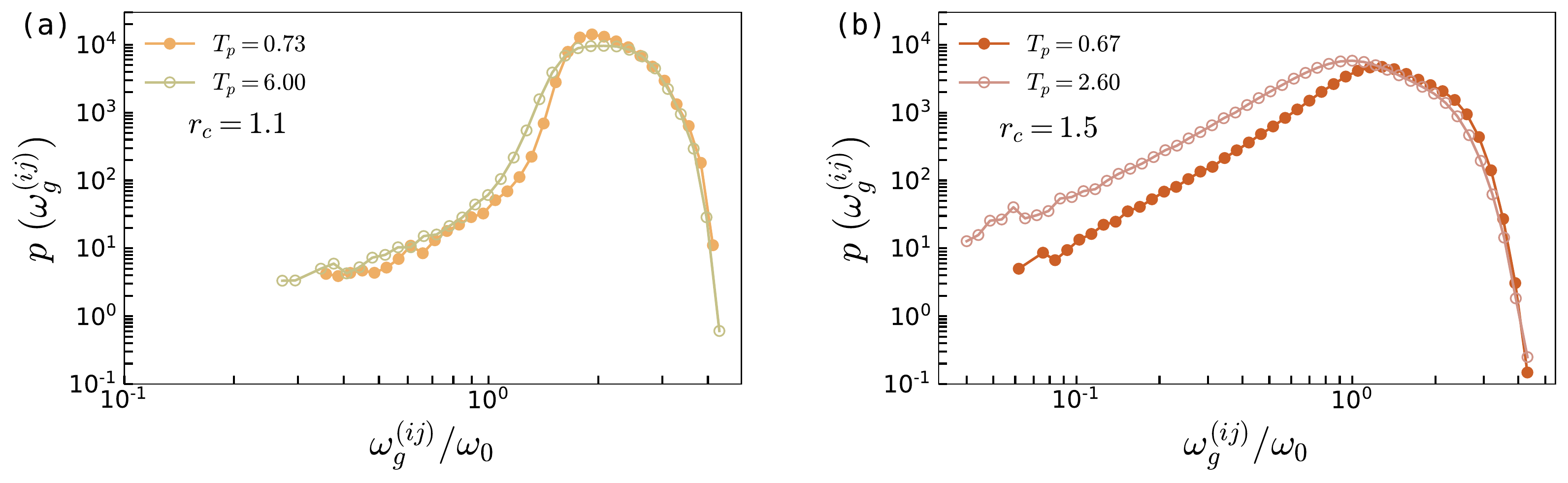}
\caption{ \footnotesize Distributions $p(\omega_g^{(ij)})$ of the dipole-response frequencies $\omega_g^{(ij)}$ (cf.~Eq.\eqref{eq:omega_g_definition}) \emph{vs.}~the dimensionless $\omega_g^{(ij)}/\omega_0$, for (a) our stickiest glasses ($r_c\!=\!1.1$) and (b) for the least sticky glasses ($r_c\!=\!1.5$), measured for their respective highest and lowest parent temperature $T_p$ as indicated by the legends. 
\label{fig:wg_distributions}}
\end{figure*}

Consistent with the behavior of other observables shown above, $A_g$ becomes a weaker function of $T_p$ as glass stickiness is increased, i.e., the glasses become thermo-mechanically inannealable. We note, importantly, that with conventional molecular dynamics methods we have employed, we are unable to supercool our model systems very deeply. In contrast, using the Swap Monte-Carlo method~\cite{LB_swap_prx}, one is able to achieve very deep supercooling of polydisperse systems, leading in some cases to a thousand-fold decrease in $A_g$, as shown in Refs.~\cite{pinching_pnas,sticky_spheres_part_1}. It is clearly important to establish in the future whether the thermo-mechanical inannealability we observe here in sticky glasses persists under deeper supercooling of their parent equilibrium states, as allowed by Swap Monte-Carlo.

\subsection{Characteristic frequency of QLMs} \label{sec:omega_gf}

The second microelastic observable we study is the aforementioned characteristic frequency $\omega_g(T_p)$ of QLMs. It has been suggested in Refs.~\cite{cge_paper,pinching_pnas,corrado_cge_statistics_jcp_2020} that the characteristic frequency of the displacement-response to locally-imposed force dipoles is a good representation of QLMs' characteristic frequency $\omega_g$. Here, we follow those suggestions; we measure 
\begin{equation}\label{eq:omega_g_definition}
    \omega_g^{(ij)} \equiv \sqrt{ \frac{\uv^{(ij)}\cdot\calBold{M}\cdot \uv^{(ij)}}{\uv^{(ij)}\cdot\uv^{(ij)}}} = \sqrt{ \frac{\dv^{(ij)}\cdot\calBold{M}^{-1}\cdot \dv^{(ij)}}{\dv^{(ij)}\cdot\calBold{M}^{-2}\cdot\dv^{(ij)}}}\,,
\end{equation}
for a large, random set of pairs of interacting particles $i,j$. Here $\uv^{(ij)}\!=\!\calBold{M}^{-1}\!\cdot\!\dv^{(ij)}$ is the displacement response to a local force dipole $\dv^{(ij)}$ imposed on the $i,j$ pair (as illustrated, e.g., in Ref.~\cite{corrado_cge_statistics_jcp_2020}), and $\calBold{M}\!=\!\frac{\partial^2U}{\partial\xv\partial\xv}$ is the Hessian matrix of the potential energy $U(\xv)$ that depends on particle coordinates $\xv$. Finally, in addition to some examples of the distributions $p(\omega_g^{(ij)}$), we also report
\begin{equation}
    \omega_g \equiv \langle\omega_g^{(ij)} \rangle_{ij}\,,
\end{equation}
where $\langle\bullet\rangle_{ij}$ denotes an average taken over interacting-pairs and glass samples. We note that in the companion paper \cite{sticky_spheres_part_1}, we followed a different route to extracting $\omega_g$; a short discussion about --- and direct comparison between --- the two methods is shown in Appendix~\ref{sec:omega_g_appendix}.

\begin{figure}
\centering
  \includegraphics[width=1.0
  \linewidth]{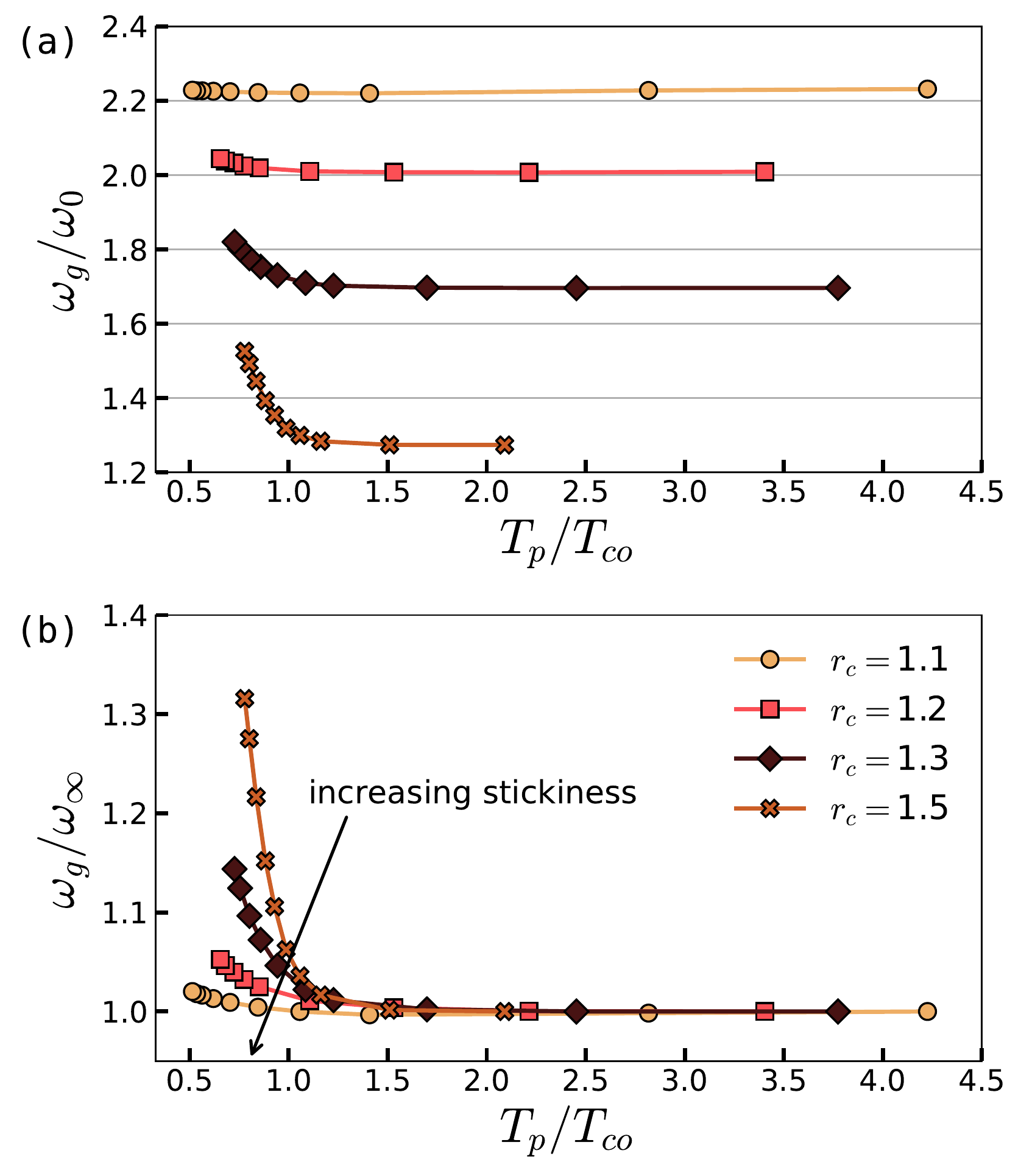}
\caption{ \footnotesize (a) Characteristic frequency $\omega_g$ of QLMs, made dimensionless by rescaling by $\omega_0\!\equiv\!c_s/a_0$, and plotted against the parent temperature $T_p$. (b) The same as (a), but rescaled by the high-$T_p$ plateau $\omega_\infty$, see text for discussion.
\label{fig:wg_vs_Tp}}
\end{figure}

In Fig.~\ref{fig:wg_distributions} we show the distributions $p(\omega_g^{(ij)})$ measured for our stickiest ($r_c\!=\!1.1$) and least sticky ($r_c\!=\!1.5$) glasses, for their lowest and highest simulated parent temperatures, as can be seen in the figure legends. Similarly to the low-frequency spectra reported in Fig.~\ref{fig:ag_mesurement}, here too we find that in the stickier glass the change in $p(\omega_g^{(ij)})$ between the highest and lowest parent temperatures $T_p$ is minor, manifesting that system's thermo-mechanical inannealability. In contrast, the distributions $p(\omega_g^{(ij)})$ for  $r_c\!=\!1.5$ glasses show a measurable difference in the amplitude of their low-frequency tails, supporting further that non-sticky glasses are thermo-mechanically annealable. We can also see that the relative width of $p(\omega_g^{(ij)})$ changes substantially between the $r_c\!=\!1.1$ and the $r_c\!=\!1.5$ glasses, indicating that increasing glass stickiness leads to micromechanical ordering. Similar trends were reported in Ref.~\cite{corrado_cge_statistics_jcp_2020}.

We next show our measurements of the means $\omega_g(T_p)$ in Fig.~\ref{fig:wg_vs_Tp}; the reported averages are taken over an immense number of interactions ($\sim\!\!10^6$), thus the data are very smooth. Here we find the same trends as seen above for the $T_p$-dependence of $p(\omega_g^{(ij)}$ and for $A_g(T_p)$: the stickier glasses show a pronounced thermo-mechanical inannealability, namely they feature almost no thermal-annealing-induced variation in $\omega_g/\omega_0$. In particular, the $r_c\!=\!1.1$ ensemble shows a total of only 3\% change in $\omega_g/\omega_0$ between its highest and lowest $T_p$'s. This should be contrasted with the variation of more than 30\% seen for the $r_c\!=\!1.5$ glasses -- a factor of at least 10 larger relative variation, compared to the total variation of $\omega_g(T_p)$ in the $r_c\!=\!1.1$ glasses, across the entire simulated $T_p$ range. 

\subsection{Localization properties of QLMs}
\label{sec:localization_properties}

It has been previously shown \cite{protocol_prerc, inst_note, pinching_pnas} that the core-size of QLMs decreases with strong supercooling of their embedding glasses' ancestral equilibrium states. Here we assess the core size of QLMs via their participation ratio $e$, defined given a mode $\psiv$ as
\begin{equation}\label{part_rat}
    e \equiv \frac{(\sum_{i} \psiv_{i} \cdot \psiv_{i})^{2}}{N\sum_{i}(\psiv_{i}\cdot \psiv_{i})^{2}}
    ,
\end{equation}
where $\psiv_{i}$ denotes the $\dbar$-dimensional vector of a mode's Cartesian components associated with the $i$th particle. One generally expects $e\!\sim\!1/N$ for (quasi)localized modes~\cite{modes_prl_2016}, and $e\!\sim\!1$ for extended modes (such as phonons). It has been shown \cite{SciPost2016,protocol_prerc,inst_note} that the participation ratio of vibrational modes at frequency $\omega$ plateaus to a value $e_0$ in the low-frequency, $\sim\!\omega^4$ scaling regime of the vDOS. Since the modes populating the scaling regime are quasilocalized, an estimation of QLMs core size --- in terms of number of particles --- is obtained via the product $Ne_0$. The extraction of $e_0$ from participation ratio data is demonstrated in Fig.~\ref{fig:Neo_example} in Appendix~\ref{sec:microscopic_observables_appendix}. 

\begin{figure}
\centering
  \includegraphics[width=1.0\linewidth]{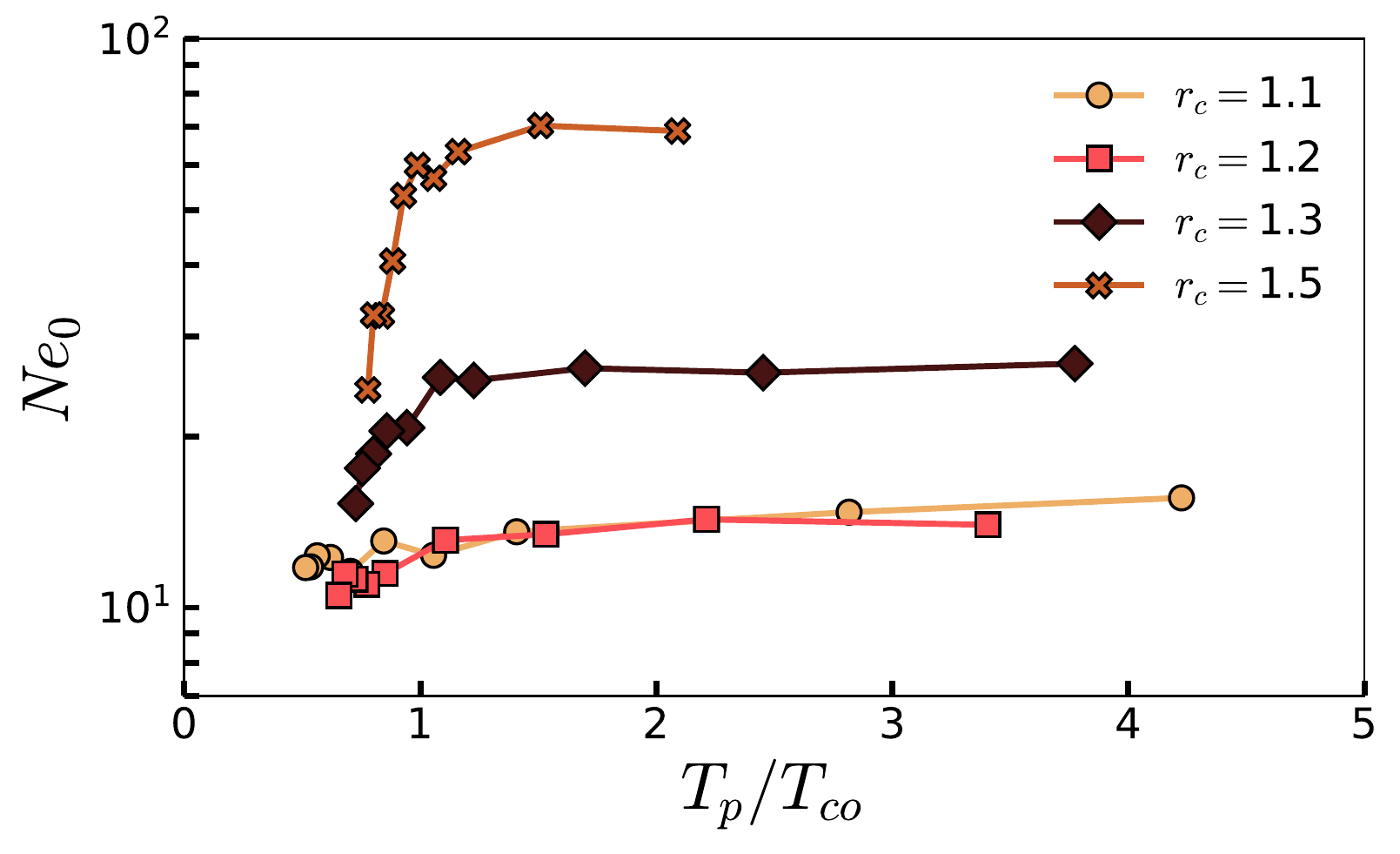}
\caption{\footnotesize Scaled mean low-frequency plateau of the participation ratio, $Ne_0$, plotted against the parent temperature $T_p$; here again we observe that increasing glass stickiness leads to thermo-mechanical inannealability. For a visual example of how $Ne_0$ is estimated, see Fig.~\ref{fig:Neo_example} in Appendix~\ref{sec:microscopic_observables_appendix}.
}\label{fig:Neo_vs_tps}
\end{figure}

Our estimations for $Ne_0$ are plotted against parent temperature $T_g$ in Fig.~\ref{fig:Neo_vs_tps}, for all $r_c$-glass-ensembles. As expected, we see that QLMs core size generally decrease with decreasing parent temperature $T_p$, consistent with the aforementioned previous observations. However, the interesting observation here is again the apparent substantial indifference of QLMs' core sizes to thermal annealing in stickier glasses, as shown above to occur for several other observables as well.

\begin{figure*}
\centering
  \includegraphics[width=1.0\linewidth]{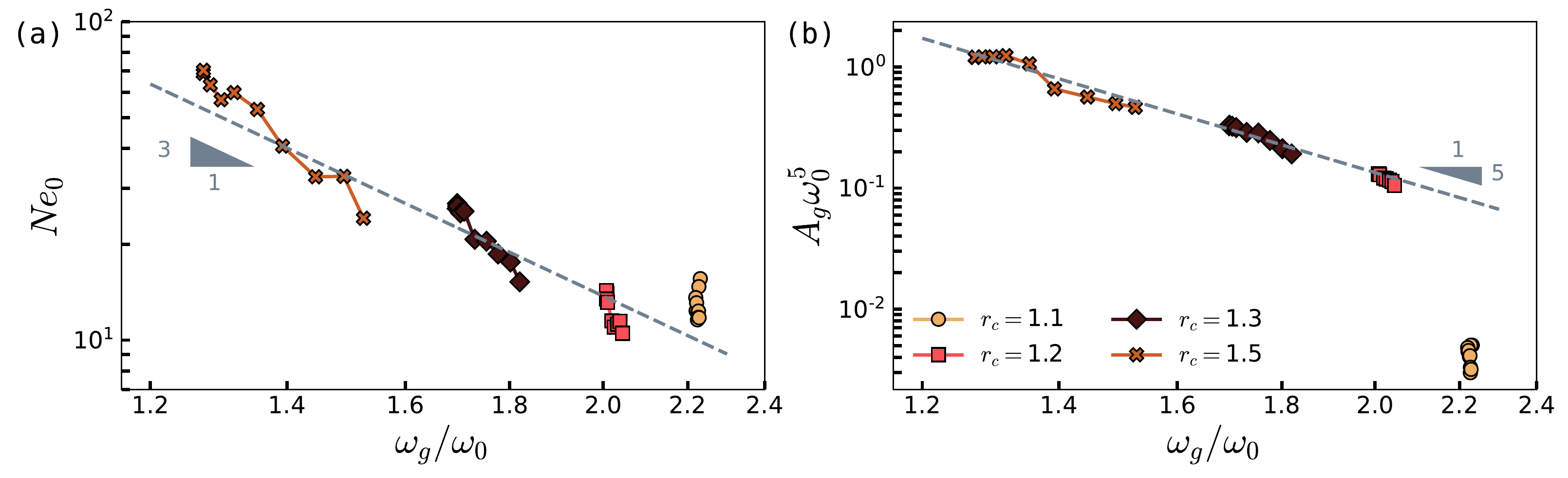}
\caption{\footnotesize (a) Scaled participation ratio $Ne_0$ plotted against the rescaled characteristic frequency $\omega_g/\omega_0$. Each point represents a different parent temperature $T_p$. The dashed line represents the relation $Ne_0\sim (\omega_g/\omega_0)^{-3}$, see text for discussion. (b) Dimensionless vDOS prefactor $A_g\omega_0^5$ plotted against the dimensionless characteristic frequency $\omega_g/\omega_0$ of QLMs, see text for discussion.
}\label{fig:Ne0_vs_wg_and_Ag_vs_wg}
\end{figure*}

\subsection{How are $A_g$, $\omega_g$ and $Ne_0$ related?}

In the preceding Subsections we have seen that the vDOS prefactor $A_g$, the characteristic frequency $\omega_g$ of QLMs, and QLMs core-size $Ne_0$, all show thermo-mechanical inannealability as glass stickiness is increased. The consistency amongst these microelastic observables supports that they are related, as previously suggested and reviewed next. 

In Ref.~\cite{pinching_pnas} it was established that the core size $\xi_g$ of QLMs is related to the latter's characteristic frequency $\omega_g$ via
\begin{equation}
    \xi_g = 2\pi c_s/\omega_g\,,
\end{equation}
where $c_s$ denotes the speed of shear waves. Since $Ne_0$ should approximately represent QLMs' core size, we expect 
\begin{equation}\label{eq:Ne0_vs_omega_g}
    N e_0 \sim (\xi_g/a_0)^3 \sim (\omega_g/\omega_0)^{-3}\,,
\end{equation}
while recalling that $a_0\!\equiv\!(V/N)^{1/3}$ is an interparticle distance, and that $\omega_0\!\equiv\!c_s/a_0$ is a characteristic frequency. Equation~\eqref{eq:Ne0_vs_omega_g} is tested in Fig.~\ref{fig:Ne0_vs_wg_and_Ag_vs_wg}a. The agreement is acceptable; the observed noise may stem from statistical errors in estimations of $e_0$ (see Appendix~\ref{sec:microscopic_observables_appendix} for details), or from a possible intrinsic error in associating the scaled participation ratio $Ne_0$ with QLMs' core size~\cite{footnote2}, or both.

We next recall that in Refs.~\cite{cge_paper,pinching_pnas} it is asserted that
\begin{equation}
    A_g\sim {\cal N}\omega_g^{-5}\,,
\end{equation}
where ${\cal N}$ represents the density per particle of QLMs. With the exception of the stickiest glasses corresponding to $r_c\!=\!1.1$, we find that $A_g$ is entirely determined by $\omega_g$, implying that ${\cal N}$ is mostly $T_p$-independent and quasiuniversal. The quasiuniversality of ${\cal N}$ was also observed and discussed in the companion paper~\cite{sticky_spheres_part_1}.

Our results are shown in Fig.~\ref{fig:Ne0_vs_wg_and_Ag_vs_wg}b; had ${\cal N}$ been universal \emph{and} $T_p$-independent, one would simply expect $A_g\!\sim\!\omega_g^{-5}$. We find that this relation approximately holds for all data points, with the very noticeable exception of the $r_c\!=\!1.1$ glasses, for which $A_g\!\ll\!\omega_g^{-5}$, as also seen and discussed in Refs.~\cite{sticky_spheres_part_1}. Thermo-mechanical inannealability is seen here as the dramatically-reduced spread of the data over the $\omega_g$ axis upon increasing glasses stickiness.


\section{Summary and outlook}
\label{sec:summary_and_discussion}

In this work we present several manifestations of a phenomenon we have coined `thermo-mechanical inannealability'. We find that, under some forms of our sticky-sphere glasses' interaction potential, a large number of micro- and macroelastic properties of glasses are seen to become largely indifferent to deeper supercooling of those glasses' ancestral equilibrium states. We demonstrated this indifference in the shear modulus $G$, as well as in several quantifiers of the statistical-mechanical properties of soft, nonphononic quasilocalized modes, including their size, characteristic frequency, and density per frequency. 

What is the degree of thermo-mechanical inannealability of common laboratory glasses? Experimental data (e.g., Refs.~\cite{Eran_mechanical_glass_transition,wang_wang_yang_wei_wen_zhao_pan_2002,experimental_inannealability_AM_2016,metallic_glass_thermomechanical_annealing_2013_wang,wang_review_2012,Li2015}) clearly indicate that macroelastic properties of metallic glasses can feature large susceptibilities to thermal annealing. In contrast, it was shown e.g., in Ref.~\cite{experimental_inannealability_AM_2016} that the shear modulus of a Ce$_{68}$Al$_{10}$Cu$_{20}$Fe$_{2}$ metallic glass changes only by about 2\% after annealing it below $T_g$ for 150 hours. Interestingly, in Ref.~\cite{experimental_inannealability_AM_2016} the same glass was also shown to be very brittle, as were (thermo-mechanically inannealable) sticky-sphere computer glasses in Ref.~\cite{itamar_brittle_to_ductile_pre_2011} -- using the same model employed here. More work is required in order to understand the extent to which the trends we have identified in simple computer glasses, as employed in this work, are relevant to laboratory glasses.

As mentioned in Sec.~\ref{sec:vdos_and_pref}, here we employed conventional molecular dynamics simulations in order to supercool our parent configurations before casting them into glassy solids. Running-time constraints do not allow one to reach very deep supercooling with these methods. It will be very interesting to observe whether the thermo-mechanical inannealability seen here persists in deeply annealed glasses made using the Swap Monte Carlo algorithm \cite{LB_swap_prx}, and to examine how it may be affected by polydispersity.

As mentioned above, here and in many other simulational work on different computer glasses (e.g., Refs.~\cite{boring_paper,LB_modes_2019}) it has been shown that $K(T_p)$ is a decreasing function of $T_p$, while it appears to be uncommon to observe a decreasing bulk modulus with thermal annealing in laboratory glasses. We speculate that this is a consequence of considering constant volume annealing, as done here, which usually leads to a decreasing bulk modulus upon thermal annealing, or constant pressure annealing, as typically done in experiments, that leads to the increase of the bulk modulus upon thermal annealing. Future research should resolve whether particular details of interaction potentials can affect the sign of $dK/dT_p$ under constant pressure annealing. 

In this work and in the companion paper~\cite{sticky_spheres_part_1} we find that tuning the interaction potential can lead to effects that resemble those of annealing on elastic properties of glasses, as also pointed out in Ref.~\cite{itamar_brittle_to_ductile_pre_2011}. Future work should carefully resolve the similarities and differences between stabilization of glasses by thermal annealing, and stabilization of glasses by tailoring the form of interaction potentials.

Finally, our findings suggest the existence of intrinsically-brittle glasses, i.e.~whose embedded defects (QLMs) are very stiff and rare, \emph{independent} of those glasses' formation history. We do not, however, find evidence for the existence of intrinsically-ductile glasses, that are both defect-rich \emph{and} thermo-mechanically inannealable. Our results suggest instead that thermo-mechanical inannealability and intrinsic brittleness share a common origin, related to glass stickiness. We leave revealing that origin to future work.

\acknowledgements

We warmly thank Srikanth Sastry, Geert Kapteijns, David Richard, Corrado Rainone, and Eran Bouchbinder for fruitful discussions. E.~L.~acknowledges support from the Netherlands Organisation for Scientific Research (NWO) (Vidi grant no.~680-47-554/3259). K.~G.~L gratefully acknowledges the computer resources provided by the Laboratorio Nacional de Superc\'omputo del Sureste de M\'exico, CONACYT member of the national laboratories network. M.~P.~C.~acknowledges support from the Singapore Ministry of Education through the Academic Research Fund MOE2017-T2-1-066 (S). Part of this work was carried out on the Dutch national e-infrastructure with the support of SURF Cooperative.

\appendix

\section{Definitions of observables}\label{sec:appendix-defs}

In this Appendix we provide precise definitions of the physical observables focused on in this study. We divide the observables to macroscopic and microscopic ones, in the next Subsections.

\subsection{Macroscopic observables}
\label{sec:macro_observables_appendix}

We start with athermal ($T\!=\!0$) elastic moduli \cite{lutsko}; the shear modulus $G$ is defined as
\begin{equation}\label{eq-G}
    G \equiv \frac{1}{V}\frac{d^2U}{d\gamma^2}= \frac{\frac{\partial^{2}U}{\partial \gamma^{2}}-\frac{\partial^{2}U}{\partial \gamma\partial \xv} \cdot \calBold{M}^{-1}\cdot \frac{\partial ^{2}U}{\partial\xv\partial \gamma}}
    {V}\,,
\end{equation}
where $\xv$ denotes particles' coordinates, $\calBold{ M}\!\equiv\!\frac{\partial^2U}{\partial\xv\partial\xv}$ is the Hessian matrix of the potential $U$, and $\gamma$ is a shear-strain parameter that parameterizes the imposed affine simple shear (in the $x$-$y$ plane) transformation of coordinates $\xv\!\to \mathBold{H}(\gamma)\cdot\xv$ with
\begin{equation}\label{shear_transformation_matrix}
\mathBold{H}(\gamma) =  \left( \begin{array}{ccc}1&\gamma&0\\0&1&0\\
0&0&1\end{array}\right)\,.
\end{equation}
We also study the nonaffine term $G_{\rm na}$ of the shear modulus, defined as
\begin{equation}
    G_{\rm na} \equiv \frac{\frac{\partial^{2}U}{\partial \gamma\partial \xv} \cdot \calBold{M}^{-1}\cdot \frac{\partial ^{2}U}{\partial\xv\partial \gamma}}{V}\,.
\end{equation}

The bulk modulus $K$ is defined as
\begin{equation}\label{eq-K}
    K \equiv -\frac{1}{\dbar}\frac{dp}{d\eta} = \frac{\frac{\partial^{2}U}{\partial \eta^{2}} - \frac{\partial U}{\partial\eta}- \frac{\partial^{2}U}{\partial \eta \partial \xv} \cdot \calBold{M}^{-1} \cdot \frac{\partial^{2}U}{\partial\xv\partial\eta}}{V\dbar}\,,
\end{equation}
where $\dbar$ is the dimension of space, $p\!\equiv\!-\frac{1}{V\dbar}\frac{dU}{d\eta}$ is the pressure, and $\eta$ is an expansive-strain parameter that parameterizes the imposed affine expansive transformation of coordinates $\xv\!\to \mathBold{H}(\eta)\cdot\xv$ as
\begin{equation}\label{dilation_transformation_matrix}
\mathBold{H}(\eta) =  \left( \begin{array}{ccc}e^\eta&0&0\\0&e^\eta&0\\0&0&e^\eta\end{array}\right)\,.
\end{equation} The Poisson's ratio $\nu$ of a glass in 3D is defined as
\begin{equation}\label{eq-poisson}   
    \nu \equiv \frac{3K-2G}{6K+2G} = \frac{3-2G/K}{6+2G/K}\,.
\end{equation}

\begin{figure}[!ht]
\centering
  \includegraphics[width=1.0\linewidth]{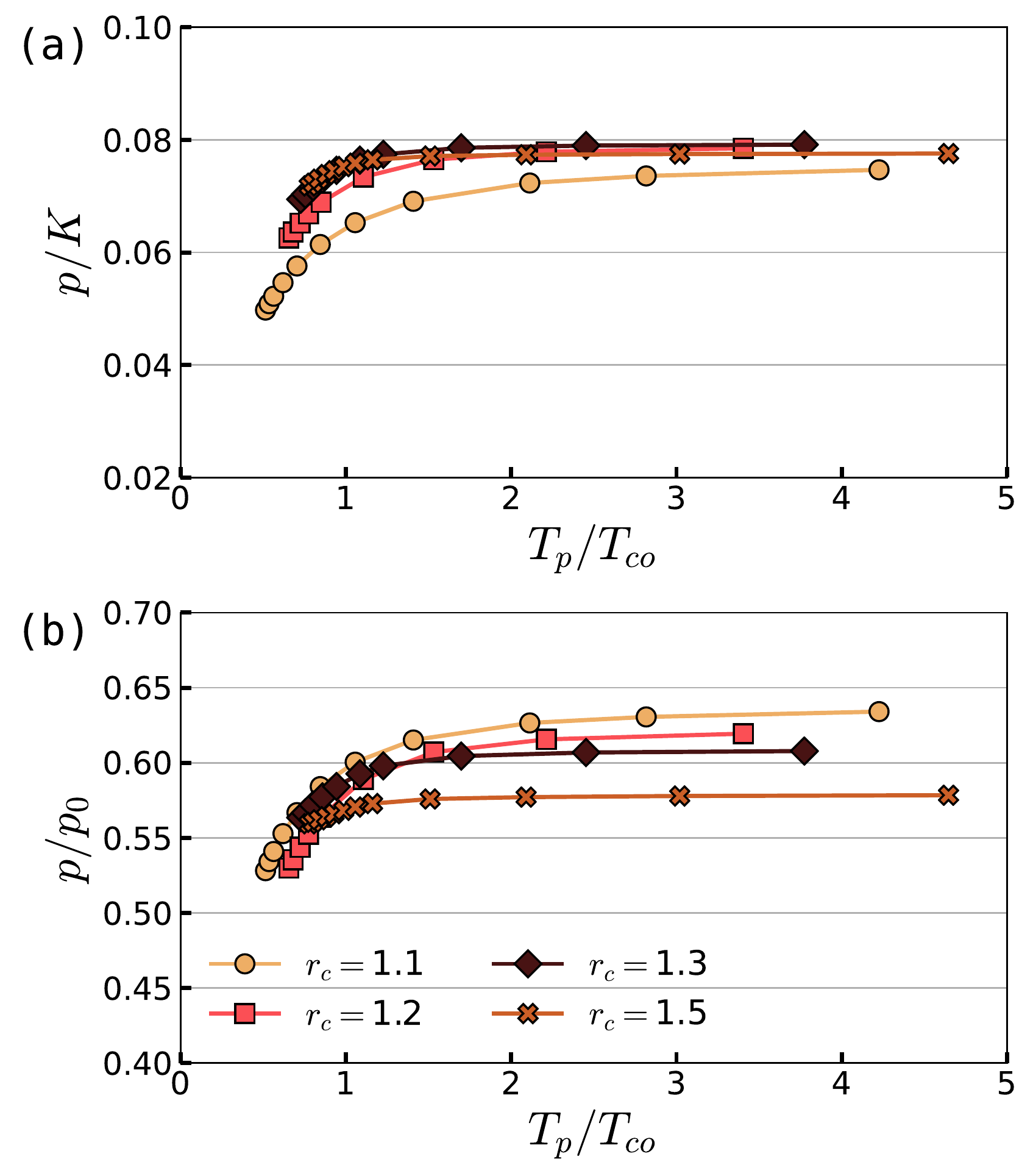}
\caption{\footnotesize We consider two dimensionless forms for our glasses' pressure: $p/K$ and $p/p_0$ (see text for the definition of $p_0$'), both indicate only minor variations between the our different $r_c$ ensembles, see further details in Appendix.~\ref{sec:macro_observables_appendix}. \label{fig:dimensionless_pressures}}
\end{figure}

In order to properly compare the pressure between different glasses, we consider here to dimensionless forms of the pressure. For the first one, we write the pressure as
\begin{equation}
    p = \frac{1}{V\dbar} \sum_{f_{ij}>0} f_{ij}r_{ij} - \frac{1}{V\dbar} \sum_{f_{ij}<0}(-f_{ij})r_{ij} \equiv p_+ - p_-\,.
\end{equation}
The above decomposition of the pressure is used to define a characteristic scale $p_0\!\equiv\!p_+ + p_-$ with respect to which the pressure can be assessed. For the second dimensionless form of the pressure, we consider the ratio $p/K$, where $K$ is the bulk modulus. The two forms of the dimensionless pressure are shown for our sticky sphere glasses in Fig.~\ref{fig:dimensionless_pressures}.

\subsection{Microscopic observables}
\label{sec:microscopic_observables_appendix}
The vibrational density of states (vDOS) is defined as
\begin{equation}
    {\cal D}(\omega) = \frac{1}{N}\bigg<\sum_\ell\delta(\omega - \omega_\ell)\bigg>\,
\end{equation}
where $\langle\bullet\rangle$ denotes an ensemble average, and $\omega_\ell$ is the vibrational frequency associated with the vibrational mode $\psiv^{(\ell)}$ that together solve the eigenvalue equation
\begin{equation}
\calBold{M}\cdot\psiv^{(\ell)} = \omega_\ell^2 \psiv^{(\ell)}\,.
\end{equation}

\begin{figure}[h!]
\centering
  \includegraphics[width=1.0\linewidth]{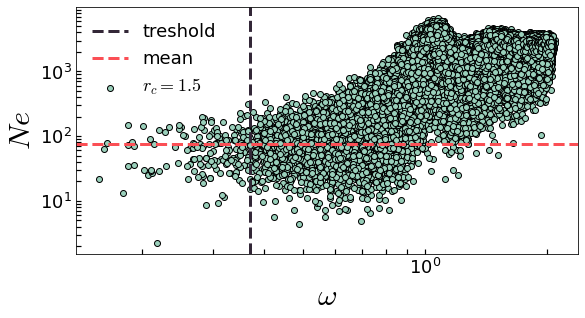}
\caption{\footnotesize Scatter plot of the scaled participation ratio $Ne$ vs.~frequency $\omega$, for our sticky-sphere glasses with $r_c\!=\!1.5$ and $T_p/T_{\mbox{\tiny co}}\!\approx\!3.0$, see text for discussion. 
\label{fig:Neo_example}}
\end{figure}

The degree of localization of vibrational modes $\psiv$ is conventionally quantified using the participation ratio $e$, defined as
\begin{equation}\label{part_rat}
    e \equiv \frac{(\sum_{i} \psiv_{i} \cdot \psiv_{i})^{2}}{N\sum_{i}(\psiv_{i}\cdot \psiv_{i})^{2}}
    ,
\end{equation}
where $\psiv_{i}$ denotes the $\dbar$-dimensional vector of a mode's Cartesian components associated with the $i$th particle. One generally expects $e\!\sim\!1/N$ for localized modes, and $e\!\sim\!1$ for extended modes. We always find that $e$ plateaus at low frequency, as shown for example in Fig.~\ref{fig:Neo_example}, and also in Refs.~\cite{SciPost2016,protocol_prerc,inst_note}.

To extract the low-frequency plateau of $e$, we set a threshold frequency --- marked by the vertical dashed line in Fig.~\ref{fig:Neo_example}, for each glass ensemble, and take the mean over all data points whose frequencies are lower than the chosen threshold, represented in Fig.~\ref{fig:Neo_example} by the horizontal dashed line. This estimation is in good agreement with the running average of $Ne_0$ at the plateau.

\section{Glass potential energy per particle}
\label{sec:potential_energy_raw_data}

\begin{figure}
\centering
  \includegraphics[width=1.0\linewidth]{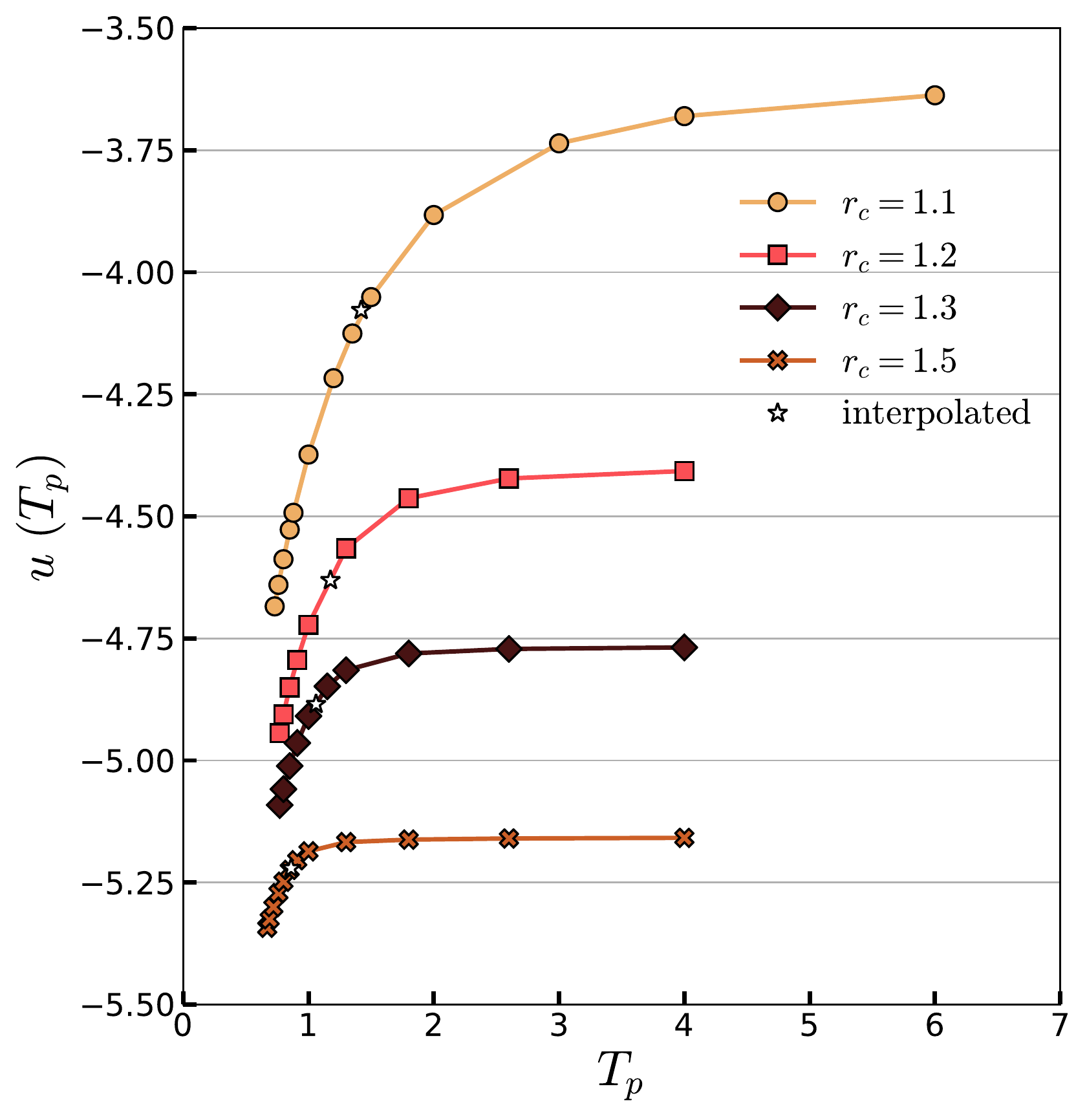}
\caption{\footnotesize Potential energy per particle, expressed in terms of \emph{simulational} units, for all studied parent temperatures $T_p$, and different glass stickiness as obtained by tuning the interaction cutoff $r_c$. The empty stars represent the interpolated energies $u(T_{\mbox{\tiny co}})$, see discussion in Sec.~\ref{sec:temperature_scale}.
\label{fig:raw_potential_energy}}
\end{figure}

In Fig.~\ref{fig:collapse_energies} of Sec.~\ref{sec:temperature_scale} we show the shifted and rescaled potential energy per particle $u(T_p)\!\equiv\!U(T_p)/N$ of sticky sphere glasses, in addition to those of other popular glass models. For readers' reference, in Fig.~\ref{fig:raw_potential_energy} we present the raw potential energy data, and the interpolated energies $u(T_{\mbox{\tiny co}})$ used in the definition of the onset temperature $T_{\mbox{\tiny on}}$, see details in Sec.~\ref{sec:temperature_scale}. 

\vspace{-1.5cm}

\vspace{0.8cm}
\section{Defining the crossover temperature via $T_p$-dependent elastic moduli}
\label{sec:alternative_Tco_appendix}

In Sec.~\ref{sec:temperature_scale} we showed how the potential energy per particle $u(T_p)$ of different glass models can be collapsed onto a master curve by a proper identification of, and rescaling by, the crossover temperatures $T_{\mbox{\tiny co}}$ for each model system. The disadvantage of this approach is that it is not useful if only one function $u(T_p)$ (for a single glass model) is available.

\begin{figure}
\centering
  \includegraphics[width=0.9\linewidth]{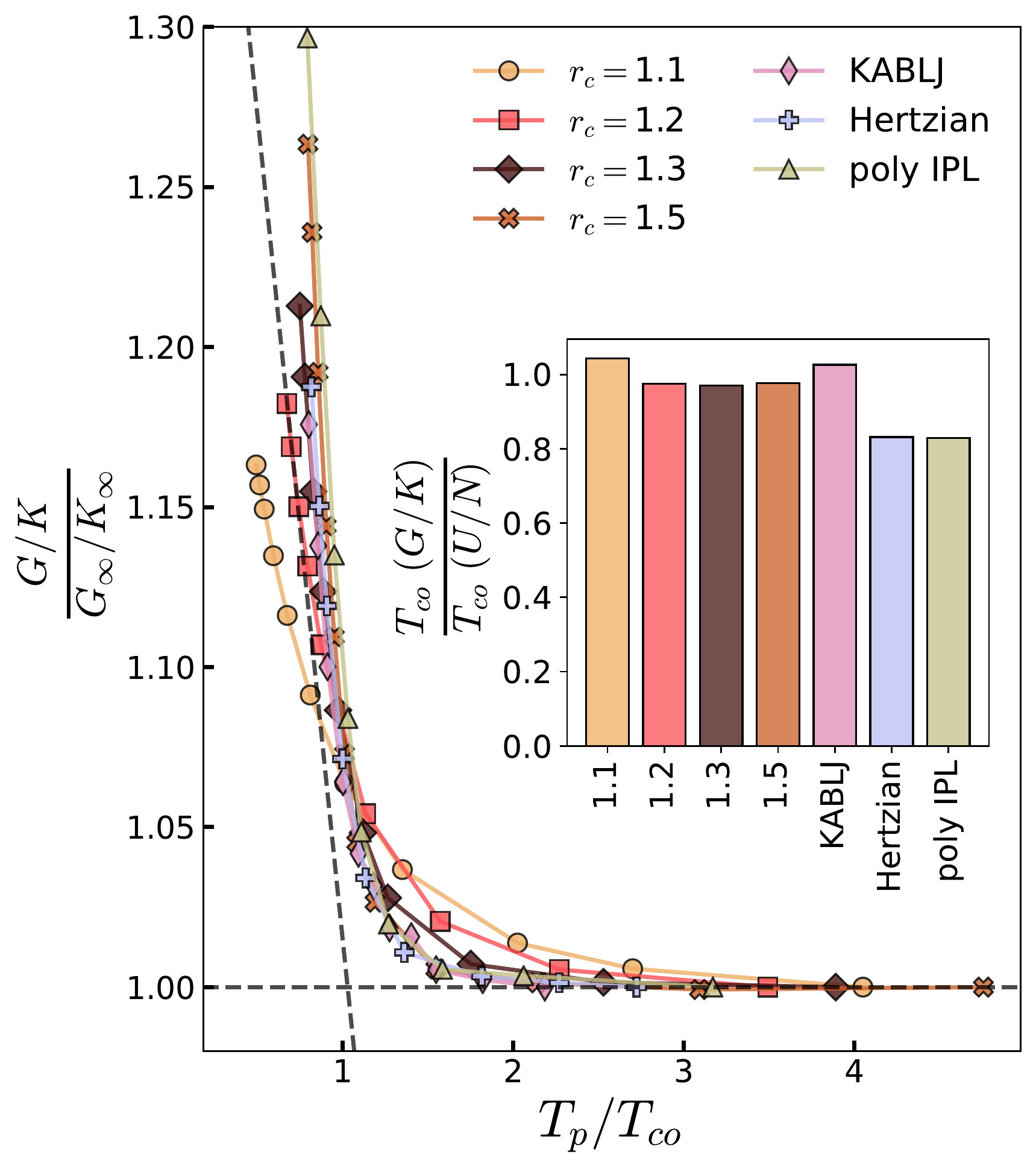}
\caption{\footnotesize Shear-to-Bulk moduli ratio $G/K$, rescaled by their high-$T_p$ limit $G_\infty/K_\infty$, and plotted against the rescaled parent temperature $T_p/T_{\mbox{\tiny co}}$. Here $T_{\mbox{\tiny co}}$ is defined as the intersection of the linearly extrapolated $G/K$ --- marked for example by the nearly-vertical dashed line --- with the high-$T_p$ limit. The inset compares between the crossover temperatures extracted as shown here, and those extracted by the $u(T_p)$ collapse shown in Fig.~\ref{fig:collapse_energies} of the main text.\label{fig:ggkk_diff_systems}}
\end{figure}

Here we offer an alternative definition, demonstrated in Fig.~\ref{fig:ggkk_diff_systems}. It amounts to constructing a linear extrapolating of the low-$T_p$ quasilinear regime of the shear to bulk moduli ratio $G/K$, towards higher $T_p$'s. An example of this extrapolation is represented in Fig.~\ref{fig:ggkk_diff_systems} by the nearly-vertical dashed line. Our alternative definition of the crossover temperature $T_{\mbox{\tiny co}}$ is given by the intersection of the extrapolated $G/K$ with the high-$T_p$ limit $G_\infty/K_\infty$. In the inset of Fig.~\ref{fig:ggkk_diff_systems} we compare the crossover temperatures $T_{\mbox{\tiny co}}$ defined via the two approaches, the one explained here and the one introduced in Sec.~\ref{sec:temperature_scale}, and find a very good agreement between the two across all models considered. We reiterate that details about the Kob-Andersen Binary Lennard-Jones (KABLJ) model, the Hertzian spheres glass, and the polydisperse inverse-power-law soft sphere glass (poly IPL) can be found in Refs.~\cite{kablj},\cite{modes_prl_2020}, and \cite{boring_paper}, respectively.  

\section{Extracting QLMs' characteristic frequency scale}
\label{sec:omega_g_appendix}

\begin{figure}
\centering
  \includegraphics[width=0.9\linewidth]{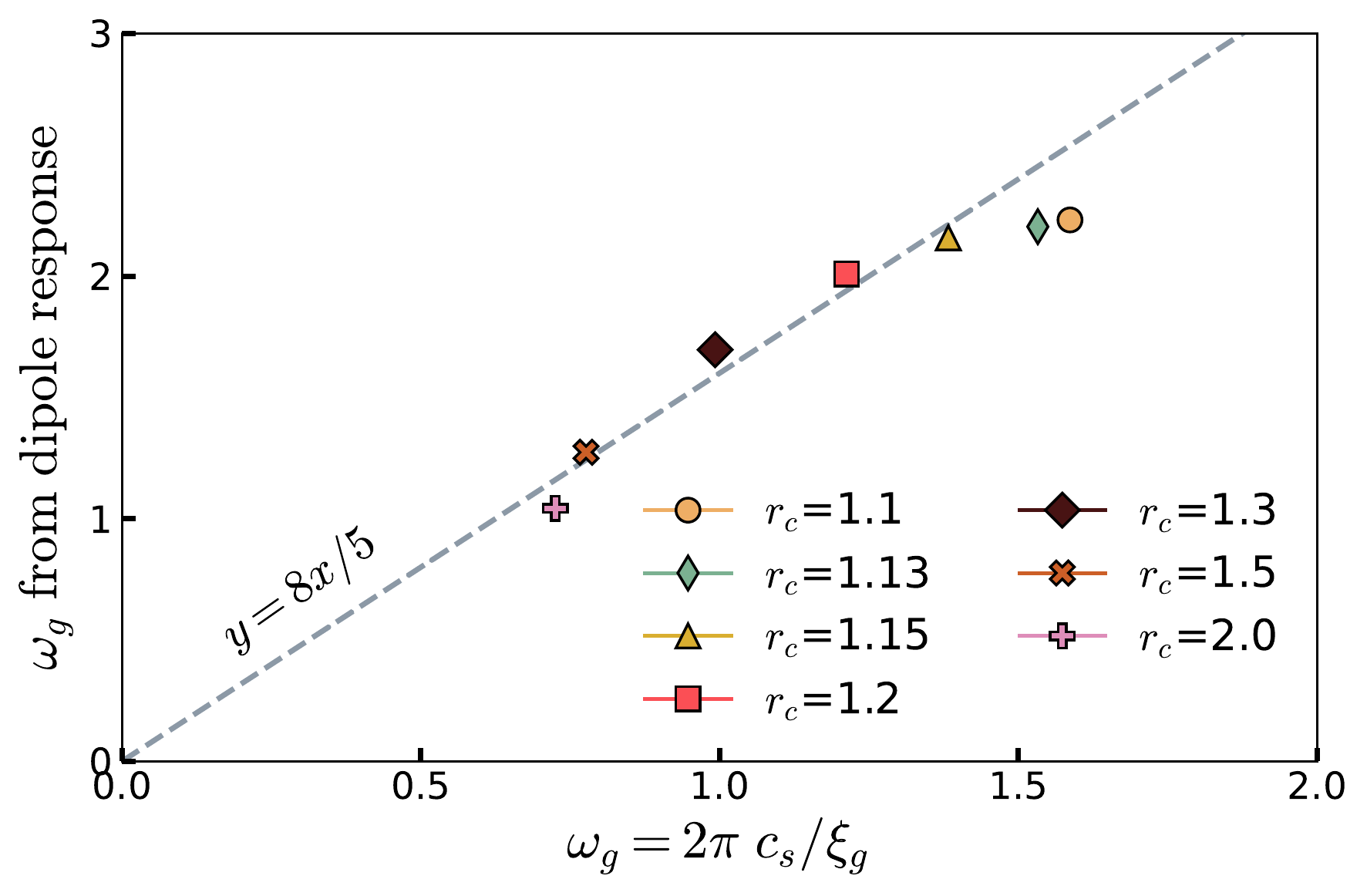}
\caption{\footnotesize Comparison of the estimation of $\omega_g$ obtained as explained in Sec.~\ref{sec:omega_gf}, to that obtained in Ref.~\cite{sticky_spheres_part_1} by following the scheme explained in this Appendix, for all $r_c$-glass ensembles (we consider the highest-$T_p$'s).
\label{fig:compare_omegag}}
\end{figure}

In the previous paper~\cite{sticky_spheres_part_1} in this series, we extracted the characteristic frequency $\omega_g$ of QLMs by (i) measuring the typical length $\xi_g$ that characterizes the response of the glass to local force dipoles, and (ii) using the relation
\begin{equation}
    \omega_g = 2\pi c_s/\xi_g \,,
\end{equation}
established in~\cite{pinching_pnas}, where $c_s$ stands for the speed of shear waves. Since this scheme requires large system sizes (we used $N\!\ge\!250$K in~\cite{sticky_spheres_part_1}), it is impractical once very long simulations are required in order to equilibrate states at very low parent temperatures. In Fig.~\ref{fig:compare_omegag} we show a good agreement between $\omega_g$ extracted via the scheme employed in this work (cf.~Sec.~\ref{sec:omega_gf}), and that extracted as explained in this Appendix.

\section{Supercooled-liquid dynamics}
\label{sec:supercooled_liquid_dynamics_appendix}


In this Appendix we report the supercooled relaxational dynamics of our different glass forming models, defined by the pair interaction cutoff $r_c$ as explained in Sec.~\ref{sec:models} of the main text. We monitor the stress autocorrelation function $c(t)$ defined as
\begin{equation}
    c(t) = V\overline{\sigma(0)\sigma(t)}/T\,,
\end{equation}
where $\overline{\bullet}$ denotes a time average, and $\sigma\!=\!V^{-1}\partial U/\partial\gamma$ with $\gamma$ denoting a shear strain parameter as defined in Appendix~\ref{sec:macro_observables_appendix}. Fig.~\ref{fig:stress_correlations} shows the stress correlations for different cutoffs $r_c$. We reiterate that a comprehensive study of the dynamical properties of our sticky sphere glasses was put forward in Ref.~\cite{Massimo_supercooled_PRL}.

\begin{widetext}
\begin{figure*}[ht!]
\centering
  \includegraphics[width=1.0\linewidth]{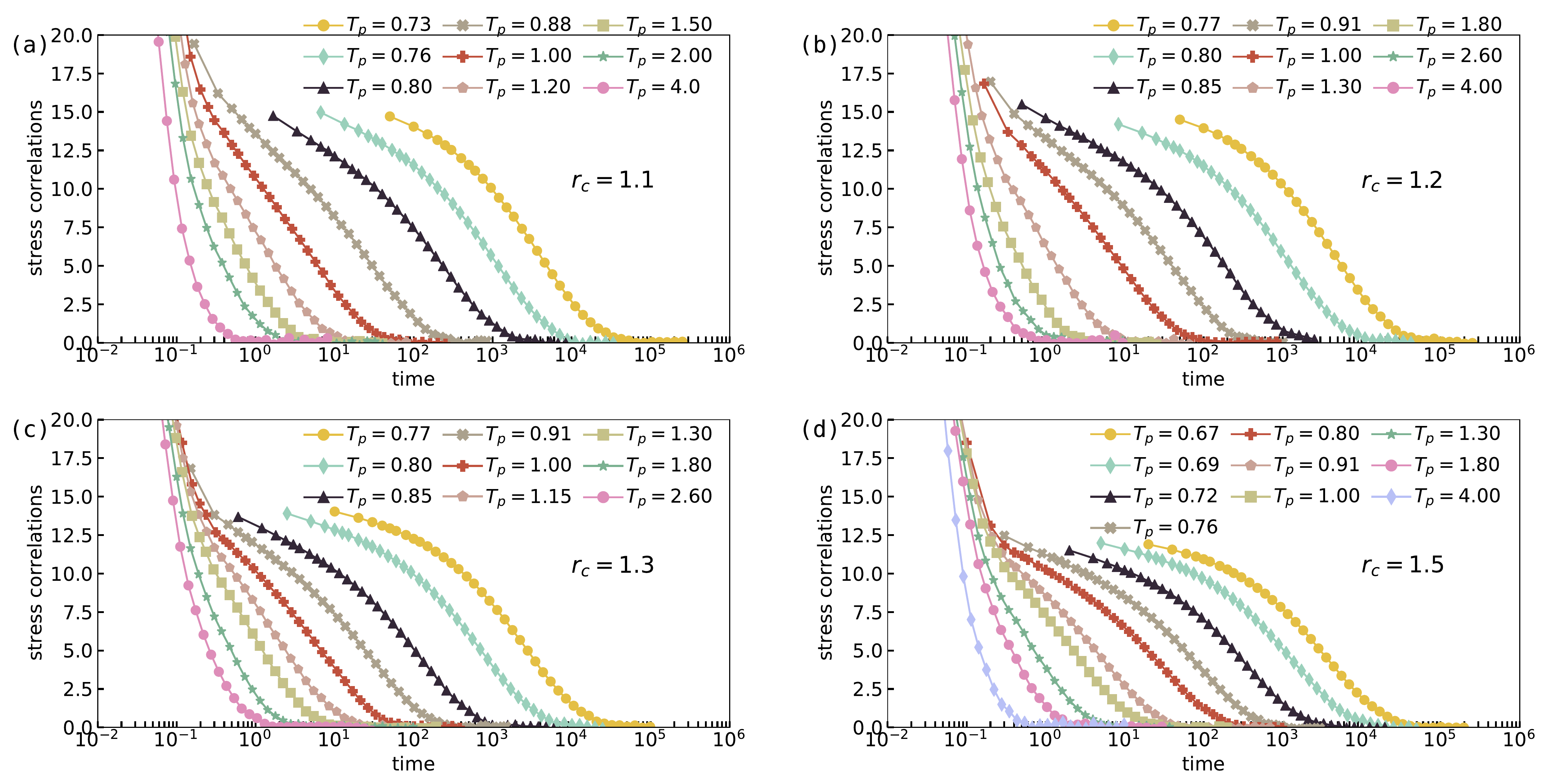}
\caption{\footnotesize Stress autocorrelation functions for our model glass formers with different pair interaction cutoffs $r_c$. 
\label{fig:stress_correlations}}
\end{figure*}  
\end{widetext}

\newpage


\begin{thebibliography}{61}
\bibitem [{foo({\natexlab{c}})}]{footnote2}%
  \BibitemOpen
  \href@noop {} {\bibfield  {journal} {\bibinfo  {journal} {In addition to
  variations in QLMs' core sizes, changes in the distribution of displacements
  within QLMs' cores could affect the participation ratio as well}\ } }\BibitemShut {NoStop}%
\end{thebibliography}
\end{document}